\documentclass[preprint,12pt]{elsarticle}

\usepackage{graphicx}

\usepackage{amssymb}
\usepackage{amsthm}

\usepackage{lineno}

\usepackage{amsmath}
\usepackage{float}
\usepackage{multirow}

\newcounter{bla}

\journal{Computer Physics Communications}

\begin{document}

\begin{frontmatter}



\title{BiconeDrag - A data processing application for the oscillating conical bob interfacial shear rheometer}


\author[a]{Pablo S\'anchez-Puga}\corref{author}
\author[a,c]{Javier Tajuelo}
\author[b]{Juan Manuel Pastor}
\author[a]{Miguel A. Rubio}\corref{author}

\cortext[author] {Corresponding author.\\\textit{E-mail address:} p.sanchez@fisfun.uned.es}
\address[a]{Departamento de F\'isica Fundamental, Universidad Nacional de Educaci\'on a Distancia, UNED, 28040 Madrid, Spain}
\address[b]{Complex System Group (ETSIAAB), Universidad Polit\'ecnica de Madrid, 28040 Madrid, Spain}
\address[c]{Departamento de F\'isica Aplicada, Universidad de Granada, 18071 Granada, Spain}

\begin{abstract}
BiconeDrag is a software package that allows one to perform a flow field based data processing of dynamic interfacial rheology data pertaining to surfactant laden air-fluid interfaces obtained by means of a rotational bicone shear rheometer. MATLAB and Python versions of the program are provided. The bicone fixture is widely used to transform a conventional bulk rotational rheometer into an interfacial shear rheometer. Typically, such systems are made of a bicone bob, which is mounted on the rheometer rotor, and a cylindrical cup. Usually, the experiment consists of measuring the response of the interface under an oscillatory stress. The program takes the values of the torque/angular displacement amplitude ratio and phase difference to compute the interfacial dynamic moduli (or complex viscosity) by consistently taking into account the  hydrodynamic flow both at the interface and the subphase. This is done by numerically solving the Navier-Stokes equations for the subphase velocity field together with the Boussinesq-Scriven boundary condition at the interface, and no slip boundary conditions elsewhere. Furthermore, the program implements a new iterative scheme devised by solving for the complex Boussinesq number in the rotor's torque balance equation. 
\end{abstract}


\end{frontmatter}



{\bf PROGRAM SUMMARY}

\begin{small}
\noindent
{\em Program Title: ``BiconeDrag" }                                        \\
{\em Program Files doi: http://dx.doi.org/10.17632/4tmy9k4ys3.1 }                                        \\
{\em Licensing provisions(please choose one): GPLv3}                                   \\
{\em Programming language: MATLAB (compatible with GNU Octave) and Python}\\
{\em Operating System: Windows, Linux and Mac OS X}
\\


{\em Nature of problem(approx. 50-250 words): Obtaining the interfacial dynamic moduli, or the complex viscosity, of a surfactant laden air-liquid interface from the experimental data obtained by means of a bicone fixture mounted on the rotor of a conventional bulk rotational rheometer. The experimental data consist on the amplitude ratio and phase difference between the torque and the angular displacement of the rotor. The coupling between the surface and subphase fluid flows require a proper representation of the hydrodynamic velocity field both at the surface and at the liquid subphase. }\\
{\em Solution method(approx. 50-250 words): We use a proper hydrodynamic model of the problem through the Navier-Stokes equations for the velocity field at the subphase, supplemented with the Boussinesq-Scriven boundary condition at the interface and no slip conditions elsewhere. The hydrodynamic equations are solved by means of a centered second order finite difference method and the flow field is used to compute the hydrodynamic drags exerted by the subphase and the interface on the bicone probe. Both calculated drags are later used in the rotor torque balance equation together with the rotor inertia term. Solving for the Boussinesq number in the torque balance equation then allows one to devise an iterative scheme that  yields improved values of the complex Boussinesq number: starting from a convenient seed one obtains a converged value of the complex Boussinesq number such that the experimental and calculated values of the torque/angle amplitude ratio coincide within a user selected tolerance. The values of the rheological variables are obtained directly from the value of the complex Boussinesq number.}\\
{\em Additional comments including Restrictions and Unusual features (approx. 50-250 words): The program is valid only for air/fluid interfaces. The interface may have or not a thin film either newtonian or viscoelastic. The subphase fluid may be newtonian or viscoelastic (having non negligible storage and loss moduli) though the user must take care of the possible frequency dependence of the dynamic moduli.}\\

\end{small}

\section{Introduction}
\label{Introduction}
Interfacial rheometers are devices that allow one to study the mechanical properties of two-dimensional fluid-fluid interfaces by relating the shear or dilatational imposed  stress with the measured deformation. Shear interfacial rheometers impose deformations that change the shape of the interface while preserving the area. Instead, dilatational rheometers change the interface area but preserve the shape of the interface. 

Usually, shear interfacial rheometers work by imposing a force on a probe floating at the interface and measuring the corresponding displacement. When the force is oscillatory with an angular frequency $\omega$, the amplitude ratio and phase difference between the oscillations of the force and the probe displacement allow one to conveniently obtain the value of the complex dynamic modulus, $G_s^* = G_s'+i G_s''$, where $G_s'$ is called the storage modulus and represents the elastic  contribution, and $G_s''$ is called the loss modulus and represents the viscous contribution. The complex interfacial viscosity is then $\eta^*_s = G_s^* /i \omega$, so that the real and imaginary parts of the complex interfacial viscosity are, respectively, $\eta_s' = G_s''/\omega$, and $\eta_s'' = G_s'/\omega$. Hence, a convenient way to characterize the interfacial viscoelasticity of fluid-fluid interfaces at a given thermodynamic state is to study the dependence of the interface response on the frequency and strain amplitude of oscillatory experiments.

Clever ways to transform bulk rotational rheometers into interfacial shear rheometers (ISRs) have been devised by substituting the regular plate-plate or cone-plate fixtures by adequate parts such as bicone bobs \cite{Erni2003} or double wall-ring (DWR) ensembles  \cite{Vandebril2010}. Among these systems DWR have in principle better resolution than bicone because of their comparatively lower inertia and smaller ratio between the perimeter in contact with the interface and the area in contact with the subphase. Indeed, that ratio is built into the non-dimensional number that governs the system performance which is the so-called complex Boussinesq number, $Bo^*$, which is usually defined as

\[ Bo^* = \frac{\eta_s^*}{L\eta},\]

\noindent where $\eta$ is the subphase bulk viscosity and $L$ is a characteristic length of the probe coming from the ratio between the length of the contact line at the probe perimeter and the contact area of the submerged part of the probe with the subphase. In the DWR $L$ is of the order of the ring radius (typically $0.5$ mm), while in the bicone is of the order of the bicone radius (typically about $30$ mm).  

In spite of their, in principle, weaker resolution in comparison with magnetic rod ISRs \cite{Brooks1999,Tajuelo2016} or with the DWR, bicone interfacial rheometers are widely used mainly to study high resistance samples. However, obtaining proper values of the dynamic moduli from the amplitude ratio and phase difference between the torque and the angular displacement in the bicone rheometer is far from trivial mainly because\cite{Tajuelo2018} i) the interface and subphase flows are coupled in a non-trivial way, and ii) the subphase is usually chosen to be a low viscosity fluid (typically water), which causes fluid inertia to come quickly into play.

Several analytical expressions have been derived that are useful for the interpretation of experimental data obtained with the bicone interfacial rheometer \cite{Oh1978,Ray1987,Nagarajan1998,Erni2003}. However, all of them pertain to the situation in which the forcing is introduced through the rotation (either steady or oscillatory) of the external cup. Unfortunately, as soon as fluid inertia comes into play (at high frequencies or low viscosities) the sheared region gets localized close to the moving part and, consequently, theoretical expressions obtained for the moving cup configuration cannot possibly represent properly the flow field of a moving bob configuration\cite{Tajuelo2018}.

Recently, Tajuelo et al. \cite{Tajuelo2018} have proposed a new and successful approach to the problem of obtaining the values of the interfacial dynamic moduli from oscillatory measurements in the interfacial bicone rheometer. The authors adapted a flow field based data processing scheme that had been successfully developed for the analogous data processing problem in the magnetic rod interfacial shear rheometer \cite{Brooks1999} either in the Helmholtz coil \cite{Reynaert2008,Verwijlen2011} or the magnetic tweezers \cite{Tajuelo2016,Tajuelo2017} configuration.

Here we make freely available a data processing software package that implements the flow field data processing scheme for dynamic interfacial rheology data obtained in air-water interfaces by means of the bicone interfacial rheometer. The software is offered in two versions, in MATLAB and Python, respectively. It uses as input data the amplitude ratio and phase difference between the imposed torque and the rotor angular displacement, which can be obtained as a part of the output data of any modern rotational rheometer, and yields proper values of the interfacial dynamic moduli or the real and imaginary parts of the complex interfacial viscosity.     

The paper is organized as follows. In Section \ref{Hydrodynamic model} we make a brief description of the system configuration and we describe the hydrodynamic model used to obtain the flow field at the subphase and the interface. A main section containing the software description follows. Finally, we have included a section devoted to present some software performance tests.  Full information concerning the implementation of the matrix of coefficients and the drag integrals are given in the appendices.
 
\section{Hydrodynamic model}
\label{Hydrodynamic model}

The measurement system is supposed to be composed of a circular cylindrical cup of radius $R_c$ and a conical bob of radius $R_b$ whose rim is placed at the air-water interface located at a height $h$ above the cup bottom surface. The total moment of inertia of the rotor+bicone assembly is $I$. We use a cylindrical coordinate system with the origin of coordinates at the point where the cylindrical symmetry axis crosses the cup bottom surface.
Moreover, nothing in the theoretical framework that we describe here precludes the possibility of the subphase itself being a viscoelastic medium. Such a case can be taken care of just by considering that the bulk subphase viscosity is a complex variable.

The minimal hydrodynamical model that takes into account the subphase-interface hydrodynamic coupling can be sketched based on the following approximations:
\begin{enumerate}[a)]
\item The velocity field at the subphase and the interface is axi-symmetric and horizontal, i.e., the velocity fields has only azimuthal component that is a function depending only on the $r$ and $z$ coordinates.
\item The interface is perfectly flat and horizontal.
\item The conical bob can be represented as a null thickness disk. This approximation can be accepted as long as the length in which the subphase velocity decays is negligible with respect to the distance from any point of the cone surface to the bottom of the measurement cell.
\end{enumerate}

In this approximation, the velocity field will have only the azimuthal component, $v_\theta (r,z,t)$, and the corresponding Navier-Stokes equation is
\begin{align}
\frac{\partial v_\theta}{\partial t}=\frac{\eta^*}{\rho}\left(\frac{\partial^2v_\theta}{\partial r^2}+\frac{\partial^2v_\theta}{\partial z^2}+\frac{1}{r}\frac{\partial v_\theta}{\partial r}-\frac{v_\theta}{r^2}\right),
\label{navier}
\end{align}
where $\rho$ is the subphase density and $\eta^*=\eta'-i\eta''$ is the complex viscosity of the subphase fluid.
 
Let us assume that the bicone performs angular oscillatory displacements with amplitude $\theta_0$ and frequency $\omega$, i.e., $\theta (t)=\theta_0\,e^{i\omega t}$. Hence, the azimuthal velocity at the bicone rim can be written as $v_{\theta,b}(t)=i\,R_b\omega\,\theta_0\,e^{i\omega t}$. In a steady oscillatory regime we can assume that the azimuthal velocity of any fluid element will be proportional to the azimuthal velocity at the bicone rim. Hence,
\begin{align}
v_\theta (r,z,t)=g^*(r,z)v_{\theta,b}(t)=i\,R_b\omega\,g^*(r,z)\theta_0e^{i\omega t},
\label{velocity}
\end{align}
where $g^*(r,z)$ is a complex amplitude function whose real and imaginary parts represent, respectively, the in-phase and out-of-phase components of the velocity field respect to the bicone motion. 
Substituting Eq. \ref{velocity} in Eq. \ref{navier} and making the spatial variables non-dimensional with the cup radius $R_c$, the Navier-Stokes equation for the fluid flow at the subphase reads 
\begin{align}
i\,Re^*\,g^*(\bar{r},\bar{z})=\frac{\partial^2g^*(\bar{r},\bar{z})}{\partial \bar{r}^2}+\frac{\partial^2g^*(\bar{r},\bar{z})}{\partial \bar{z}^2}+\frac{1}{\bar{r}}\frac{\partial g^*(\bar{r},\bar{z})}{\partial \bar{r}}-\frac{g^*(\bar{r},\bar{z})}{\bar{r}^2},
\label{navier_g}
\end{align}

\noindent where $Re^* = \frac{\rho \omega R_c^2}{\eta^*}$ is the complex Reynolds number, and the overbar indicates non-dimensional variables . 

The boundary conditions are no slip at the cup walls and bottom, zero velocity along the cylindrical symmetry axis, and no slip at the bicone submerged surface. In terms of the amplitude function $g^*$, these boundary conditions can be represented as
\begin{align}
&g^*(\bar{r},0)=g^*(1,\bar{z})=0,\nonumber\\
&g^*(0,\bar{z})=0,\nonumber\\
&g^*(\bar{r}\leq \bar{R}_b,\bar{h})=\frac{\bar{r}}{\bar{R}_b}.
\label{boundary1}
\end{align}
At the air water interface, i.e., for $\bar{R}_b<\bar{r}<1$ and $\bar{z}=\bar{h}$, we use the Boussinesq-Scriven boundary condition \cite{Reynaert2008,Scriven1960}, which involves the complex interfacial viscosity. In cylindrical coordinates it reads
\begin{align}
\frac{\partial g^*}{\partial \bar{z}}=Bo^*\frac{\partial}{\partial \bar{r}}\left(\frac{1}{\bar{r}}\frac{\partial}{\partial \bar{r}}\left(\bar{r}\,g^*\right)\right)\mathrm{\, ,\, at \,\,\,}\bar{R}_b<\bar{r}<1 ,\, \bar{z}=\bar{h}.
\label{boundary-bouss}
\end{align}
where now the complex Boussinesq number is defined as
\[ Bo^* = \frac{\eta_s^*}{R_c\eta^*},\]

In order to have a full description of the problem, the hydrodynamic equations have to be supplemented with the equation for the rotor dynamics which is the torque balance equation:
\begin{align}
T^*(t)+T_{\mathrm{sub}}^*(t)+T_{\mathrm{surf}}^*(t)=I\frac{\partial^2\theta(t)}{\partial t^2},
\label{balance}
\end{align}
where $T^*$ is the applied torque and $T_{\mathrm{sub}}^*$ and $T_{\mathrm{surf}}^*$ are, respectively, the subphase and surface drag terms (asterisks are used to indicate complex variables).
The torque imposed on the rotor+conical bob ensemble and its angular displacement, which can be represented as
\begin{align}
\theta^*(t)&=\theta_0e^{i\omega t},\nonumber\\
T^*(t)&=T_0e^{i\omega t-\delta}=T_0e^{-i\delta}e^{i\omega t} =T_0^*e^{i\omega t},
\end{align}
\noindent allow us to define the complex amplitude ratio between the torque and the angular displacement, $AR^*$, as
\begin{align}
AR^*&=\frac{T_0^*}{\theta_0},\nonumber\\
\delta&=\mathrm{arg}\left(AR^*\right).
\end{align}

The expressions for the surface and subphase drags are 
\begin{align}
&T_{\mathrm{sub}}^*=-i\omega2\pi R_b\eta^*\theta_0e^{i\omega t}\int_0^{R_b}r^2\left.\left(\frac{\partial g^*}{\partial z}\right)\right\vert_{z=h}dr,\nonumber\\
&T_{\mathrm{surf}}^*=i\omega2\pi R_b^2R_c\,Bo^*\eta^*\theta_0e^{i\omega t}\left( R_b\left.\left(\frac{\partial g^*}{\partial r}\right)\right\vert_{r=R_b,\,z=h}-1\right).
\label{drags}
\end{align}
Taking into account the Eqs. \ref{velocity}, \ref{balance}, and \ref{drags}, the amplitude ratio, the azimuthal velocity amplitude  function $g^*$, and the complex Boussinesq number, $Bo^*$, are related by the following expression

\begin{align}
AR^* =& i\omega 2 \pi R_b\eta^* \left[\int_0^{R_b}r^2 \left(\frac{\partial g^*}{\partial z}\right) \Bigg|_{z=h}dr \right. \nonumber \\
& \left. -R_b R_c\, Bo^*\left( R_b \left(\frac{\partial g^*}{\partial r}\right) \Bigg|_{r=R_b,\,z=h}-1\right)\right]-I\omega^2.
\label{AR-eq}
\end{align}

Interestingly, $g^*$ depends on $Bo^*$ through the Navier-Stokes equation and the Boussinesq-Scriven boundary condition and, solving for the complex Boussinesq number at the torque balance equation \ref{AR-eq}, $Bo^*$ in its turn depends on $g^*$ through the hydrodynamic drag terms, 
\begin{align}
Bo^*=\frac{-AR^*-I\omega^2+i\omega2\pi R_b\eta^*\int_0^{R_b}r^2\left.\left(\frac{\partial g^*}{\partial z}\right)\right\vert_{z=h}dr}{i\omega2\pi \eta^*R_b^2R_c\,\left( R_b\left.\left(\frac{\partial g^*}{\partial r}\right)\right\vert_{r=R_b,\,z=h}-1\right)}.
\label{Bo-eq}
\end{align}
The structure of the problem is highly non-trivial and, consequently, it cannot be solved directly. However, the very same structure of the problem makes it well suited to be solved through an iterative numerical scheme in a simple way: first, a seed value is given to $Bo^*$, second, the Navier-Stokes equation is numerically solved and $g^*$ is found, third, the numerically obtained velocity amplitude function $g^*$ is used to calculate the hydrodynamic drag terms in equation \ref{Bo-eq}, that together with the experimentally measured value of the complex amplitude ratio, $AR^*_{\mathrm{exp}}$, yield a new value for the complex Boussinesq number; finally, this new value is reintroduced in the Navier-Stokes equation and the loop is iterated till convergence, within a given tolerance, occurs. A detailed description of the implementation of the iterative process is given in subsection \ref{Sec:IterativeProcess}.

\section{Software description}
\label{Software Overview}

\subsection{Software overview}
The program basically reads the data corresponding to the experimental torque-angular displacement complex amplitude ratio, $AR^*_{\mathrm{exp}}$, from an input data file written in a spreadsheet format which is located in the specified file path and applies the iterative process to each data line in a while-loop structure. For each input data file, the program creates an output file made containing output data lines giving the calculated values of the dynamic surface moduli, $G'$ and $G''$, and the frequency at which the experiment was ran.

We provide full package versions for, both, MATLAB and Python environments.  MATLAB is one of the most widely used and well known high level languages in the field of scientific computing, while Python is getting established as a reference programming language in the free software community. Usually, MATLAB codes are considered to be slow in comparison to optimized and compiled C, C++, and Fortran codes because, in its basic form, it is a code interpreter. However, by taking some simple programming precautions, such as avoiding the excessive use of for-loops, casting the problem into a linear algebra form with sparse matrices, and/or using pre-compiled functions, a MATLAB code can be rendered rather efficient. The code has been originally written in the MATLAB (R2018a) environment but we have checked its compatibility with the GNU Octave environment obtaining successful results. The MATLAB (Octave) program version has been checked in OS X and Linux operating system without problems.

Alternatively, we have also developed a version of the code written in Python in order to offer another free option to process the data experiments. Python is usually included in modern Linux distributions. Tests with Python in such systems have also been satisfactory. 

\subsection{Parameter and Data Input}

Some model parameters must be set in order for the computations to work. The list of those parameters is shown in Table \ref{table:Tbl_Param}. To facilitate the general view we have classified the parameters according to different aspects of the problem. The way by which the parameter values are set is through a script. An example of a typical script is shown in \ref{section:exampleScript}. 

Among the parameters required by the program, there are some geometrical and dynamical parameters of the rheometer, and the physical parameters of the subphase to be set. All of them are defined in Table \ref{table:Tbl_Param} although some of them deserve particular comments.

For instance, the parameter denoted by the variable name $b$ is the coefficient of frictional torque of the rheometer. This parameter may enter into play when measuring weak resistance interfaces. The way to measure it was discussed in the Supporting information of Ref. \cite{Tajuelo2018}. 

In agreement with our previous remark, all along the program, the subphase viscosity is a complex variable. Hence, viscoelastic subphases can be accounted for by just inserting the value (known in advance) of the complex bulk subphase viscosity corresponding to the frequency at which the interfacial rheology measurements were made.

The number of subintervals in the radial and vertical coordinates ($N$ and $M$, respectively), control the spatial resolution in the computation, the convergence tolerance (name tolMin), that is used to decide whether convergence has occurred, and the maximum number of iterations allowed (iteMax) set an upper limit to the iteration process to preclude the system iterating without limit in case convergence is not achieved.

The experimental data are fed by means of an input data file having a spreadsheet format. The control software of most commercial rotational rheometers directly provides such an output data file which typically has an user selected column structure. The program looks for all files at the current path having a $``\_exp.txt"$ end pattern in alphabetical order. Hence, it is mandatory to change the input data file name for it to have such a pattern. Then the user must put into the program the numbers of the columns that contain, respectively, the frequency at which the measurement was made (colIndexFreq), the amplitude of the torque/angular displacement ratio (coIndexAR), and the phase lag of the angular displacement behind the torque (colIndexDelta).   

The output data file consist of an output data line corresponding to each input data line. The name of the output data file is the same of the experimental data file, finished in $``\_out.txt"$ instead of $``\_exp.txt"$. Each line will contain the corresponding values of the frequency, the interfacial dynamic moduli ($G_s'$ and $G_s''$), the real and imaginary parts of the complex interfacial viscosity ($\eta_s'$ and $\eta_s''$), the real and imaginary parts of the Boussinesq number ($Bo'$ and $Bo''$), the modulus and argument of the converged torque/angle amplitude ratio ($\lvert AR^*_{calc}\rvert$, $\arg (AR^*_{calc})$), the elapsed time in each iterative process and the number of iterations until convergence, all parameters on units of the international system. The paths of both, the input and output data files, are given through the script.

\begin{table}[H]
\begin{tabular}{ |l|l|l|l| }
\hline
& & & \\
Aspect & Name & Units & Concept \\
& & & \\
\hline
\multirow{4}{*}{Geometry} & \multirow{2}{*}{h} & \multirow{2}{*}{m} & Vertical distance between the interface \\  
 & & &  and the bottom of the cup \\
 & $R_b$ & m & Bicone radius  \\ 
 & $R_c$ & m & Cup radius \\ \hline
\multirow{4}{*}{Dynamics} & \multirow{2}{*}{inertia} & \multirow{2}{*}{kg$\cdot$m$^2$} & Moment of inertia of the rotor + bicone \\
& & & assembly\\
& \multirow{2}{*}{b} & \multirow{2}{*}{kg$\cdot$m$^2\cdot $s$^{-1}$} & Coefficient of the frictional torque \\
& & & of the rheometer \\ \hline
\multirow{2}{*}{Subphase} & rho$\_$bulk & kg$\cdot$m$^{-3}$ & Density of the subphase \\
& \multirow{1}{*}{eta$\_$bulk} & \multirow{1}{*}{Pa$\cdot$s} & Complex subphase viscosity \\ \hline
\multirow{4}{*}{Mesh} & \multirow {2}{*}{N} & \multirow {2}{*}{} & Number of subintervals in the radial \\
 & & & coordinate $r$ \\
 & \multirow {2}{*}{M} & \multirow {2}{*}{} & Number of subintervals in the vertical \\
 & & & coordinate $z$ \\
\hline
\multirow{2}{*}{Iteration} & iteMax & & Maximum number of iterations allowed \\
 & tolMin & & Convergence tolerance \\ \hline
\multirow{11}{*}{Input/output data} & \multirow {3}{*}{colIndexAR} & \multirow {3}{*}{} & Ordinal number of the data file column \\ 
& & & that contains the modulus of the\\
& & & amplitude ratio \\
& \multirow {3}{*}{colIndexDelta} & \multirow {3}{*}{} & Ordinal number of the data file column \\ 
& & & that contains the phase of the\\
& & & amplitude ratio \\
& \multirow {3}{*}{colIndexFreq} & \multirow {3}{*}{} & Ordinal number of the data file column \\ 
& & & that contains the frequency of the\\
& & & oscillations \\ 
& \multirow {1}{*}{inputFilepath} & \multirow {1}{*}{} & Path to the Input file \\ 
& \multirow {1}{*}{outputFilepath} & \multirow {1}{*}{} & Path to the Output file \\ \hline
\end{tabular}
\caption{Program parameters.}
\label{table:Tbl_Param}
\end{table}

\subsection{General flowchart}
\begin{figure}[H]
\centering
\includegraphics[width=.7\linewidth]{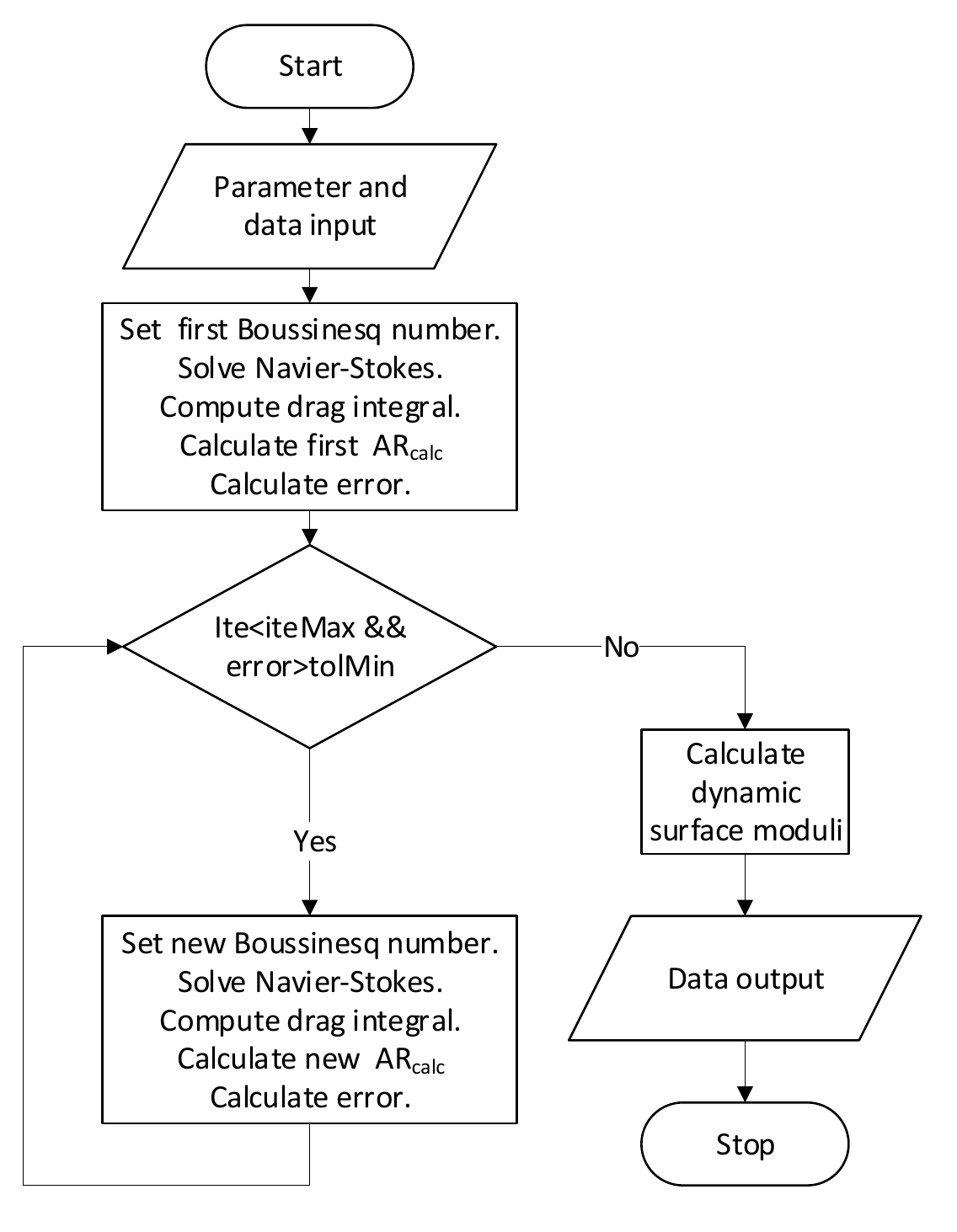}
\caption{General Flowchart}
\label{fig:Flowchart}
\end{figure}

According to the above given description the general flowchart of the program is simple and can be seen in Fig. \ref{fig:Flowchart}. First, the required parameter values are input. The next item in the flowchart contains several tasks that have to be executed in the following order: i) obtain the Boussinesq number seed, ii) solve the Navier-Stokes equation with the boundary conditions, iii) calculate the hydrodynamic drag terms, and iv) obtain the calculated complex torque/angle amplitude ratio.

Then the value of $AR^*_{calc}$ is checked against $AR^*_{exp}$ for the tolerance value set. If the tolerance criterion is failed  steps ii) to iv) are repeated until convergence is found. If the tolerance criterion is fulfilled the values of the rheological properties are calculated and a new line is written in the output data file.

\subsection{Hydrodynamic calculations}
\label{Data Analysis}

We take advantage of the symmetry of the problem, so that it suffices to solve it in the rectangle defined by $0\le \bar{r} \le 1$ and $0 \le \bar{z} \le \bar{h}$. Then, the Navier-Stokes equation, Eq. \ref{navier_g}, with the boundary conditions shown in equations \ref{boundary1} and \ref{boundary-bouss} is solved by means of a second order centered finite differences method. 

\begin{figure}[H]
\centering
\includegraphics[width=.7\linewidth]{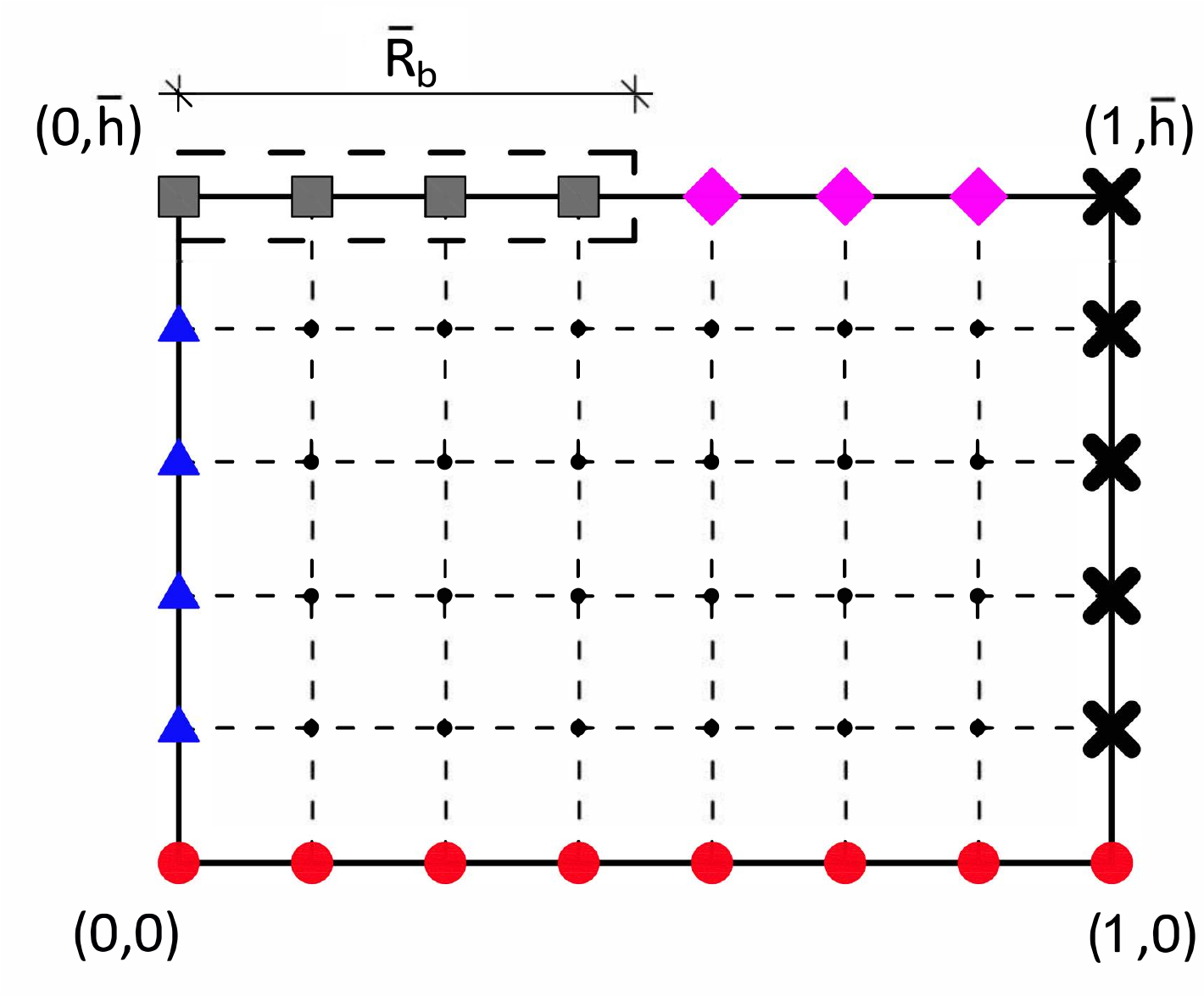}
\caption{Sketch of the mesh and boundaries in a meridian half plane with the cylindrical symmetry axis at the left side and the cup wall at the right side. Red circles: Cup bottom; black crosses: Cup lateral wall; blue triangles: Symmetry axis; gray squares: Bicone bob surface; magenta diamonds: Interface.}
\label{fig:Mesh}
\end{figure}

The mesh consists of $N$ and $M$ evenly spaced subinterval in the $\bar{r}$ and $\bar{z}$ coordinates, respectively, with the coordinate system origin located at the center of the cup bottom. A cartoon version of the mesh is illustrated in Fig. \ref{fig:Mesh}. Red circles and black crosses  indicate nodes located at the bottom of the cup and the cup lateral wall, respectively. Blue triangles are nodes located at the symmetry axis of the system, gray squares represent the nodes located at the bicone surface, and the magenta diamonds represent the nodes located at the interface. The values of $M$ and $N$ can be set at will. The usual geometrical configuration for bicone systems involves cup radius about $R_c = 4$ cm, and bicone to cup bottom distances about $h=2$ cm, hence, in the following we have worked with an $N$ to $M$ ratio of $2$. 

In the above-mentioned approximation, the flow field is fully described by the value of the complex function $g^*(\bar{r},\bar{z})$ at the mesh nodes. Hence,  in the mesh we will have 
\begin{align}
g_{j,k}^*=g^*\left( (j-1)\frac{1}{N},(k-1)\frac{\bar{h}}{M} \right)=g^*(\bar{r}, \bar{z}),\hspace{0.5cm}\\
\forall j,k\in\mathbb{Z}\, \Vert \, 1\leq j\leq N+1\, , \, 1\leq k\leq M+1,
\end{align}
where $j$ and $k$ represent the node coordinates. In such a mesh and within a centered second order finite differences scheme, the fluid flow equations at each mesh node can be written in terms of the values of $g_{j,k}^*$ at the four nearest neighbours (three for the nodes at the boundaries). 
For the internal nodes (nodes not at a boundary) the discrete Navier-Stokes equation, at node $(j,k)$, takes the form (see the appendices for details) 
\begin{align}
i\,Re^*\,g_{j,k}^*=&N^2\left( g_{j+1,k}^*+g_{j-1,k}^*-2g_{j,k}^*+\frac{g_{j+1,k}^*-g_{j-1,k}^*}{2(j-1)}-\frac{g_{j,k}^*}{(j-1)^2} \right)\nonumber\\
&+\left( \frac{M}{\bar{h}} \right)^2\left( g_{j,k-1}^*+g_{j,k+1}^*-2g_{j,k}^* \right),\nonumber\\
&\forall j,k\in\mathbb{Z}\, / \, 2\leq j\leq N,\,2\leq k\leq M.\label{Eq:navier-nodos}
\end{align}

Now, if we rearrange the values of $g_{j,k}^*$ as a column vector $g^*_\alpha$, of size $(N+1)(M+1)$, with 
\begin{align}
g^*_\alpha=g_{j,k}^*, \; \; \mathrm{with} \; \; \alpha = (k-1)(N+1)+j,\\
\forall j,k\in\mathbb{Z}\, \Vert \, 1\leq j\leq N+1\, , \, 1\leq k\leq M+1,\nonumber \\
\forall \alpha \in\mathbb{Z}\, \Vert \, 1\leq \alpha\leq (N+1)(M+1), \nonumber
\end{align}

\noindent we can write the problem as a linear equations system 
\begin{equation}
\textbf{A}\cdot\textbf{g}=\textbf{b},
\label{eq:lin_equation_syst}
\end{equation}
\noindent where $\textbf{g} = g^*_\alpha$, $\textbf{A}$ is  the coefficients matrix (a square sparse matrix of size $(N+1)(M+1) \times(N+1)(M+1)$, and $\textbf{b}$ is the independent terms vector (a null vector except for the nodes at the bicone bob interface). We refer the reader to the appendices for details. 

Note that these expressions are valid for Matlab code, where indexes $j$ and $k$ run from $1$ to $N+1$ and $1$ to $M+1$, respectively. In the Python code the expressions are slightly different because index $j$ and $k$ run from $0$ to $N$ and from $0$ to $M$. Finally, the integral in the bicone drag term appearing in the torque balance equation is calculated by means of the compound trapezoidal rule.

\subsection{Iterative process}
\label{Sec:IterativeProcess}
The iterative process is implemented by starting from a seed for $Bo^*$ and subsequently repeating the following steps until convergence occurs:
\begin{enumerate}
\item Solving the Navier-Stokes equation, Eq. \ref{navier_g}, with the boundary conditions given by Eqs. \ref{boundary1}, and \ref{boundary-bouss} , i.e., numerically solving the linear equations system \ref{eq:lin_equation_syst}.
\item Computing the surface and subphase hydrodynamic torques out of the solution of the Navier-Stokes equation, Eq. \ref{drags}, i.e., applying Eq.~\ref{eq:numer_integr}.
\item Checking whether the iteration scheme has converged or not. When convergence is achieved the iterative process ends.
\item Obtaining a new value of $Bo^*$ through the torque balance equation including the experimental value of the complex torque/angular displacement amplitude ratio Eq. \ref{Bo-eq}, and repeat steps 1 to 3. 
\end{enumerate}

At variance with respect to the usual procedure \cite{Vandebril2010,Verwijlen2011,Tajuelo2016,Tajuelo2018}, we prefer not to focus on the convergence of $Bo^*$, but on finding the value of the complex Boussinesq number that gives a complex torque/angle amplitude ratio, $AR^*_{calc}$, that is the closest possible to the experimental value $AR_{exp}^*$. Hence, we solve for $Bo^*$ in Eq. \ref{Bo-eq}, and use the expression

\begin{align}
Bo^{*\{i+1\}}=\frac{-AR_{exp}^*-I\omega^2+i\omega2\pi R_b\eta^*\int_0^{R_b}r^2\left.\left(\frac{\partial g^{*\{i\}}}{\partial z}\right)\right\vert_{z=h}dr}{i\omega2\pi \eta^*R_b^2R_c\,\left( R_b\left.\left(\frac{\partial g^{*\{i\}}}{\partial r}\right)\right\vert_{r=R_b,\,z=h}-1\right)}
\label{Bo-iter}
\end{align}

\noindent to obtain a new corrected value $Bo^{*\{i+1\}}$.
After convergence occurs, the dynamic surface moduli, $G_s'$ and $G_s''$ are obtained from the definition of the complex Boussinesq number
\begin{align}
G_s^*=\omega R_c\eta^* Bo^*.
\label{G_s}
\end{align}

Typically, the number of iterations required to reach convergence depends on the relative importance of the surface and subphase drags and the rotor inertia. The higher the surface drag, the less iterations are needed. For the reported experiments, the number of iterations varies from 1 to 10.

Choosing an appropriate seed is important. In our experience, good results are usually obtained by choosing as starting value of $Bo^*$ the one corresponding to the ideal solution having a linear velocity profile (constant shear rate) across the interface.

\begin{align}
Bo^{*\{i=0\}}=\frac{R_c-R_b}{i2\pi\omega\eta^* R_b^2R_c^2}\left(AR^*_{exp}-AR^*_{clean}\right)
\end{align}
\noindent where
\begin{align}
AR^*_{clean}=\frac{i\pi\omega R_b^4\eta^*}{2h}-I\omega^2
\end{align}

The unprocessed experimental data given by  the rheometer correspond to the torque/angle amplitude ratio, therefore, we have chosen to implement a convergence condition based on $AR_{calc}^*$ being close to $AR_{exp}^*$ as follows:

\begin{align}
\Bigg|\frac{(AR_{pp}^*)_{calc}^{\{i\}}-(AR_{pp}^*)_{exp}}{(AR_{pp}^*)_{exp}}\Bigg|\leq tolMin
\label{Eq:tol_condition}
\end{align}

Hence, the convergence condition always refers to the experimental complex amplitude ratio from the input data files.

\subsection{Software package structure}

The software package consists of a main script, where every program parameters  are input (see Table \ref{table:Tbl_Param}), and several functions.
The data processing is started through a script that reads the program parameters and calls the main function $postprocessingBiconeCPC.m$ which in its turn calls two subroutines:
\begin{itemize}
\item $GetFilenames.m$: It returns a $1\times n$ cell array with the $n$ experiment filenames at the selected input file path.
\item $solve\_NS\_bicono.m$: It solves the hydrodynamic equations with the adequate boundary conditions and returns a column vector, $g_{\alpha}^*$, with $(N+1)\cdot(M+1)$ elements containing the values of $g^*(\bar{r},\bar{z})$. Full details are given in  \ref{FD-details}.
\end{itemize}

After obtaining a new iterated value $Bo^{*\{i+1\}}$ the main function tests the convergence condition (Eq.\ref{Eq:tol_condition}) and repeats the iterative process until the convergence condition is fulfilled.

\section{Program performance}
\label{Program performance}
In this section we illustrate the performance of the software package on the different tests we have made. We report on the results of typical test on the main parts of the program: the Navier-Stokes solution, the convergence of the iterative process, the reproducibility of the results through its performance working on synthesized data, and the program performance when working with noisy experimental data. 

\subsection{The solution of the Navier-Stokes equation}

To our knowledge there is not an analytical solution for the cup and bicone bob problem in the oscillatory bob configuration so that we do not have the possibility to check the fluid flow fields obtained against exact solutions. However, as was shown in Ref. \cite{Tajuelo2018}, simple changes on the boundary conditions in equations \ref{boundary1} allow us to cast the problem in other cup-bicone bob problems with different combinations of standing, oscillating, or steadily motions of either part. Thorough checks against the analytical solutions in the literature \cite{Oh1978} were reported in Ref. \cite{Tajuelo2018} obtaining an excellent agreement.

\begin{figure}[H]
  \begin{minipage}{0.5\textwidth}
  \centering
  \includegraphics[width=\linewidth]{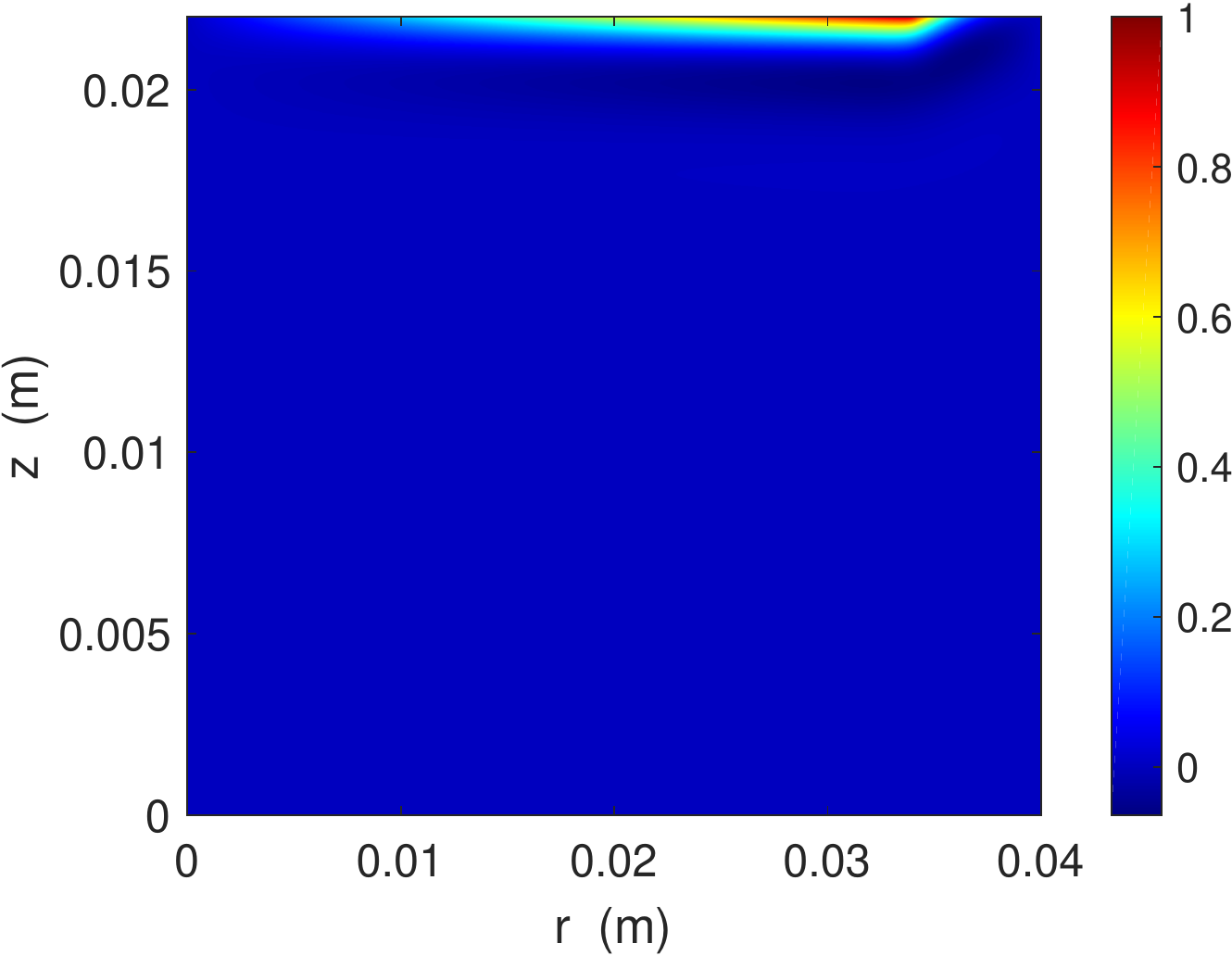}
  \end{minipage}
  \begin{minipage}{0.5\textwidth}
  \centering
  \includegraphics[width=\linewidth]{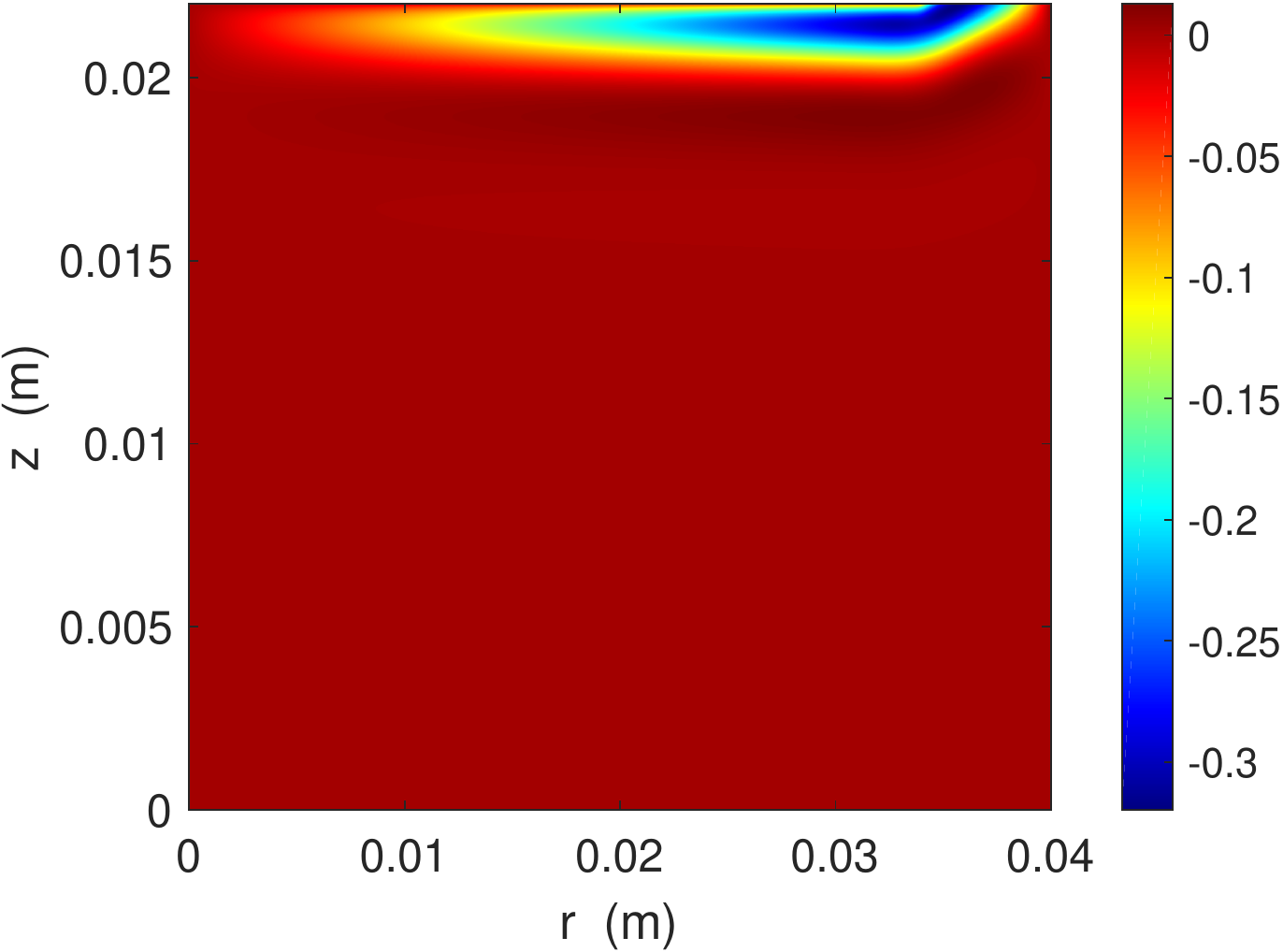}
  \end{minipage}
\caption{Color coded plots of (a) $\Re[g^*(r,z)]$, and (b) $\Im[g^*(r,z)]$ at $Bo^*=0.1-0.1i$.}
\label{fig:ColorMap}  
\end{figure}

For the case of interest here, i.e., the oscillatory bicone configuration,  we have thoroughly analyzed the calculated flow fields in a large variety of situations. As an illustrative example we show in Fig. \ref{fig:ColorMap}	color coded graphs of the real (a) and imaginary (b) parts of the azimuthal velocity amplitude function, $g^*(r,z)$, for $Bo^* = 0.1 - 0.1i$, obtained with a $2520\times1260$ mesh with no image smoothing. At such a low value of $Bo^*$, fluid inertia effects are expected to appear and strong velocity gradients can be appreciated close to the moving part, i.e., the bicone surface. Note that the values of the real part of the velocity are larger in absolute value than those corresponding to the imaginary part, which are negative. 

\begin{figure}[H]
  \begin{minipage}{0.45\textwidth}
  \centering
  \includegraphics[width=\linewidth]{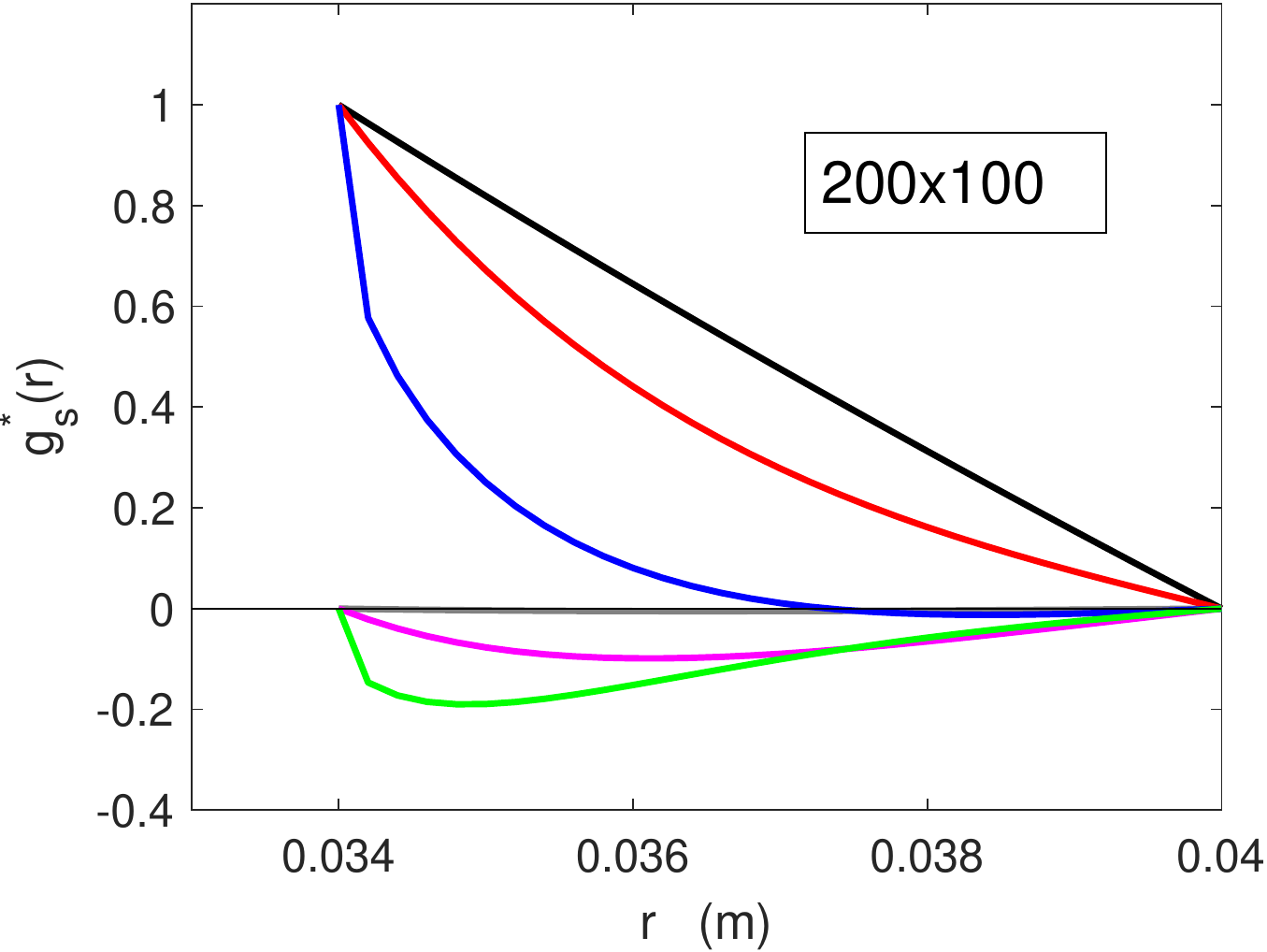}
  \end{minipage}
  \begin{minipage}{0.45\textwidth}
  \centering
  \includegraphics[width=\linewidth]{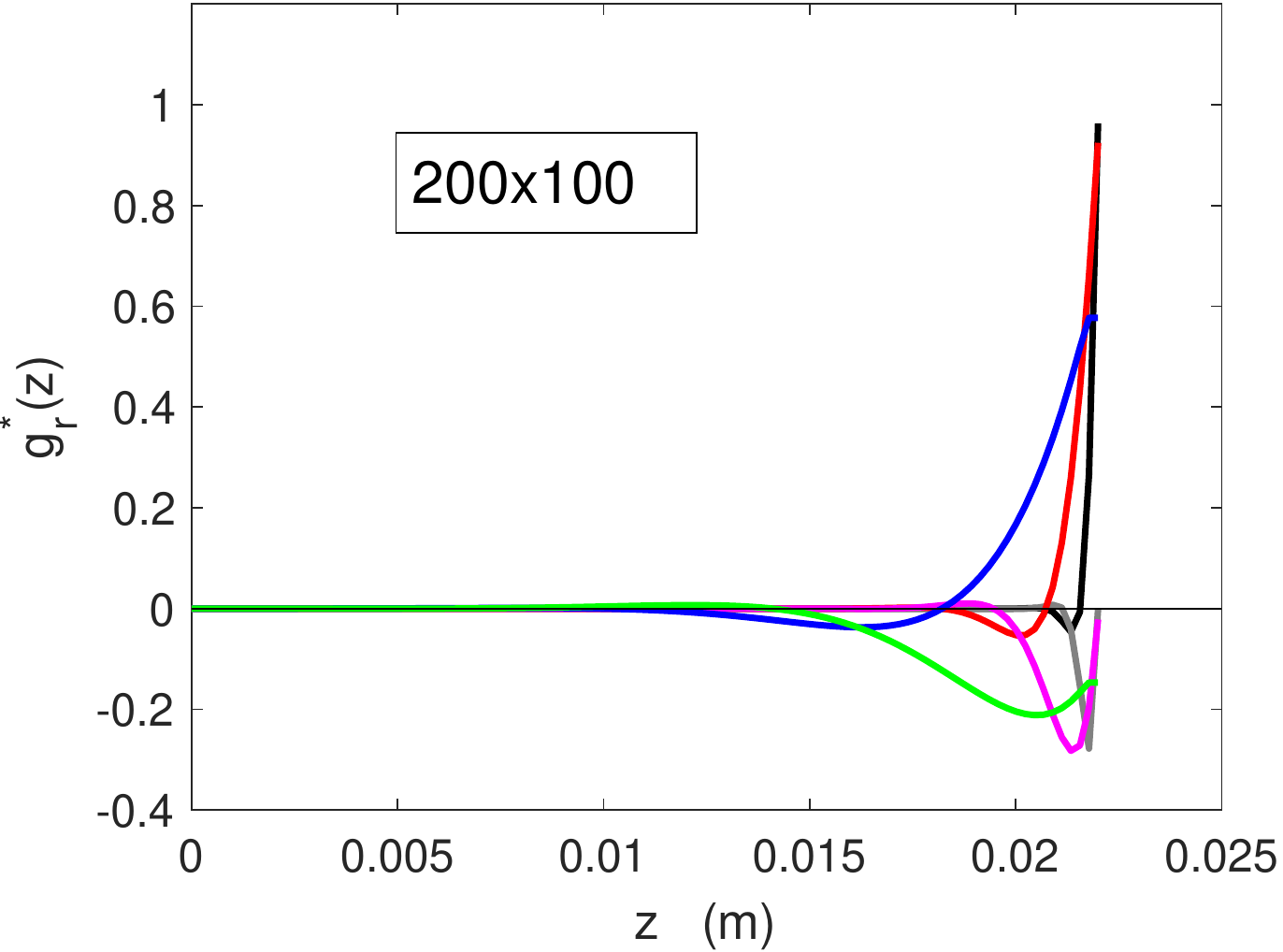}
  \end{minipage}
  
  \begin{minipage}{0.45\textwidth}
  \centering
  \includegraphics[width=\linewidth]{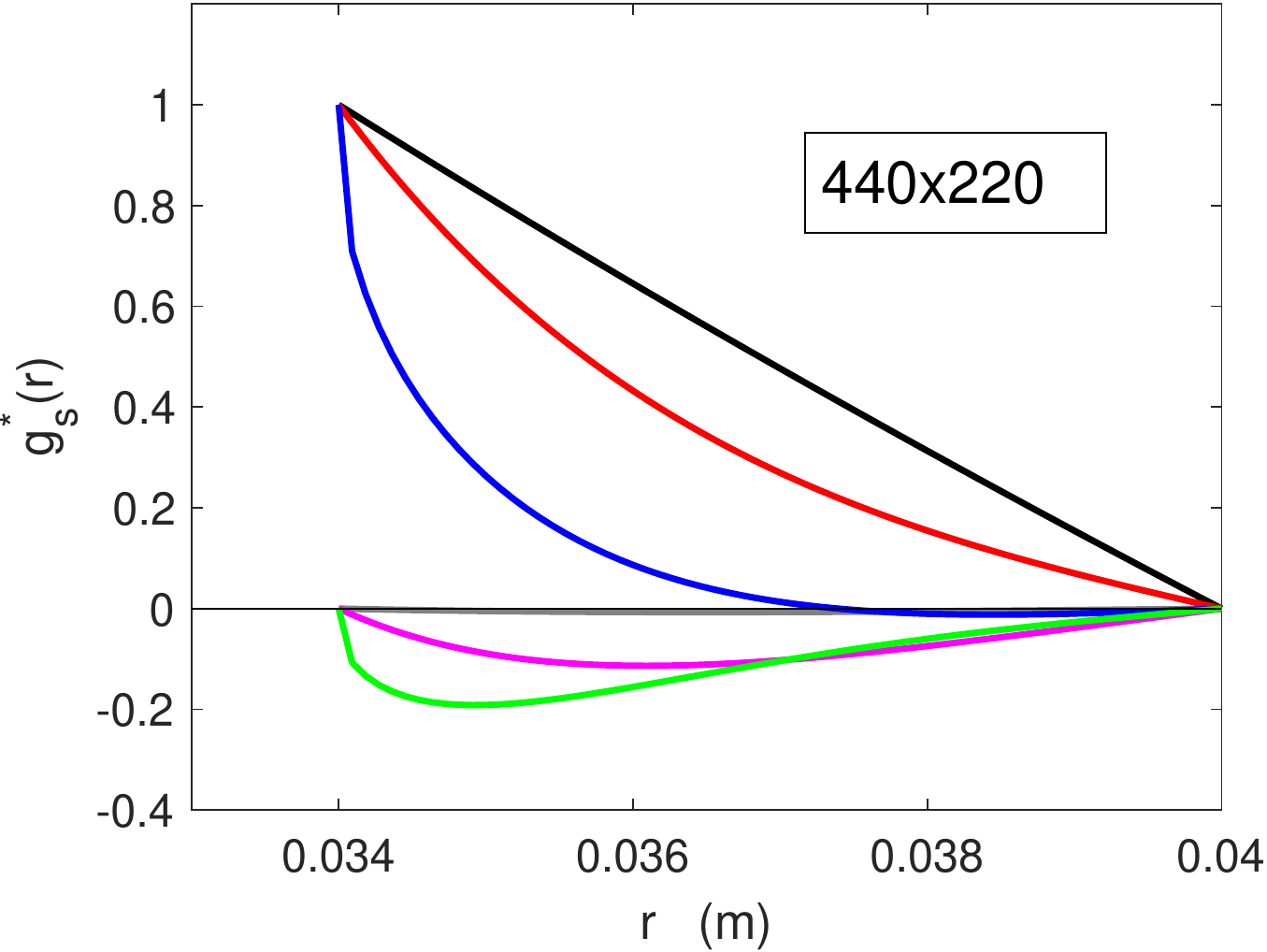}
  \end{minipage}
  \begin{minipage}{0.45\textwidth}
  \centering
  \includegraphics[width=\linewidth]{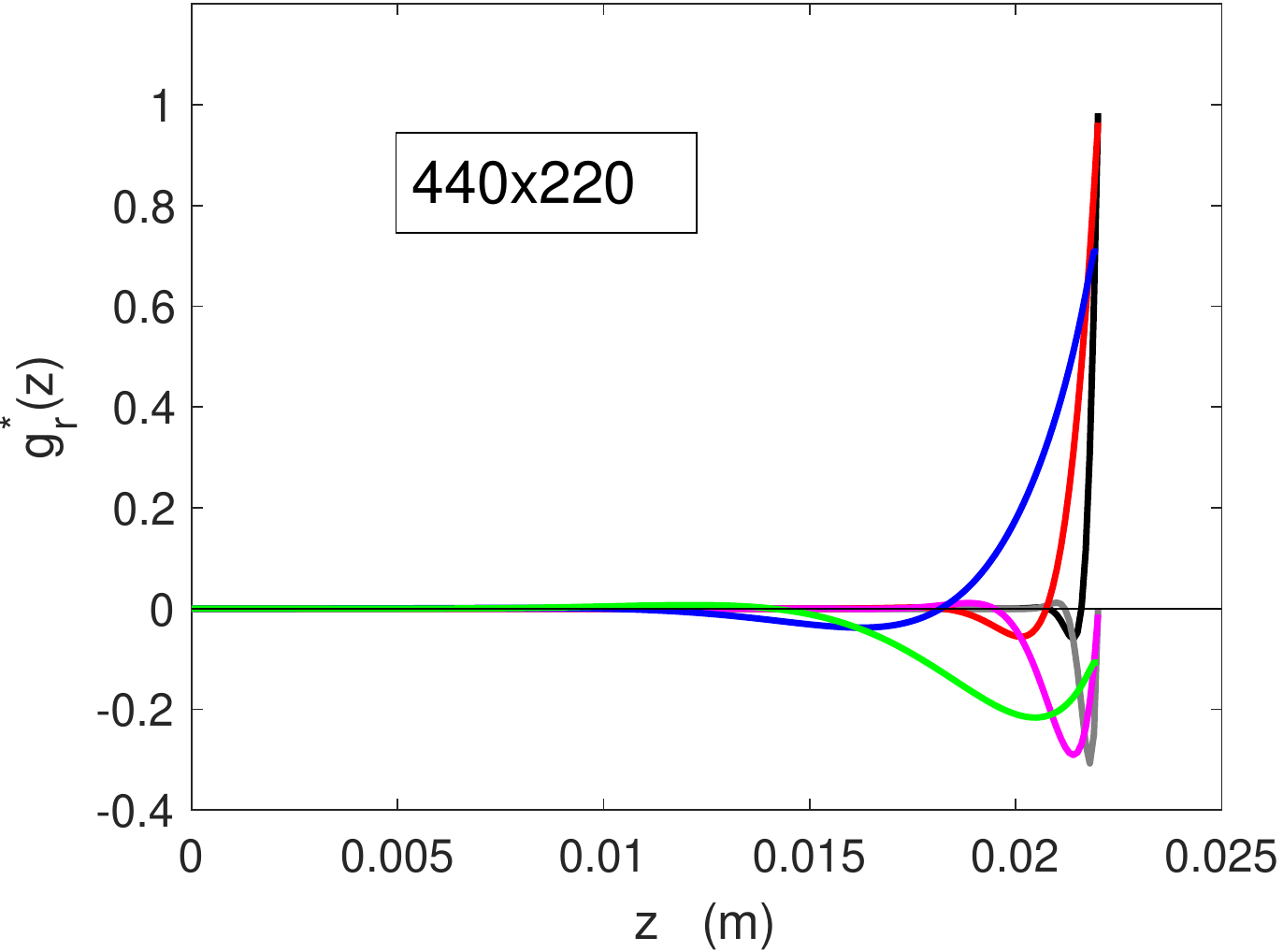}
  \end{minipage}

  \begin{minipage}{0.45\textwidth}
  \centering
  \includegraphics[width=\linewidth]{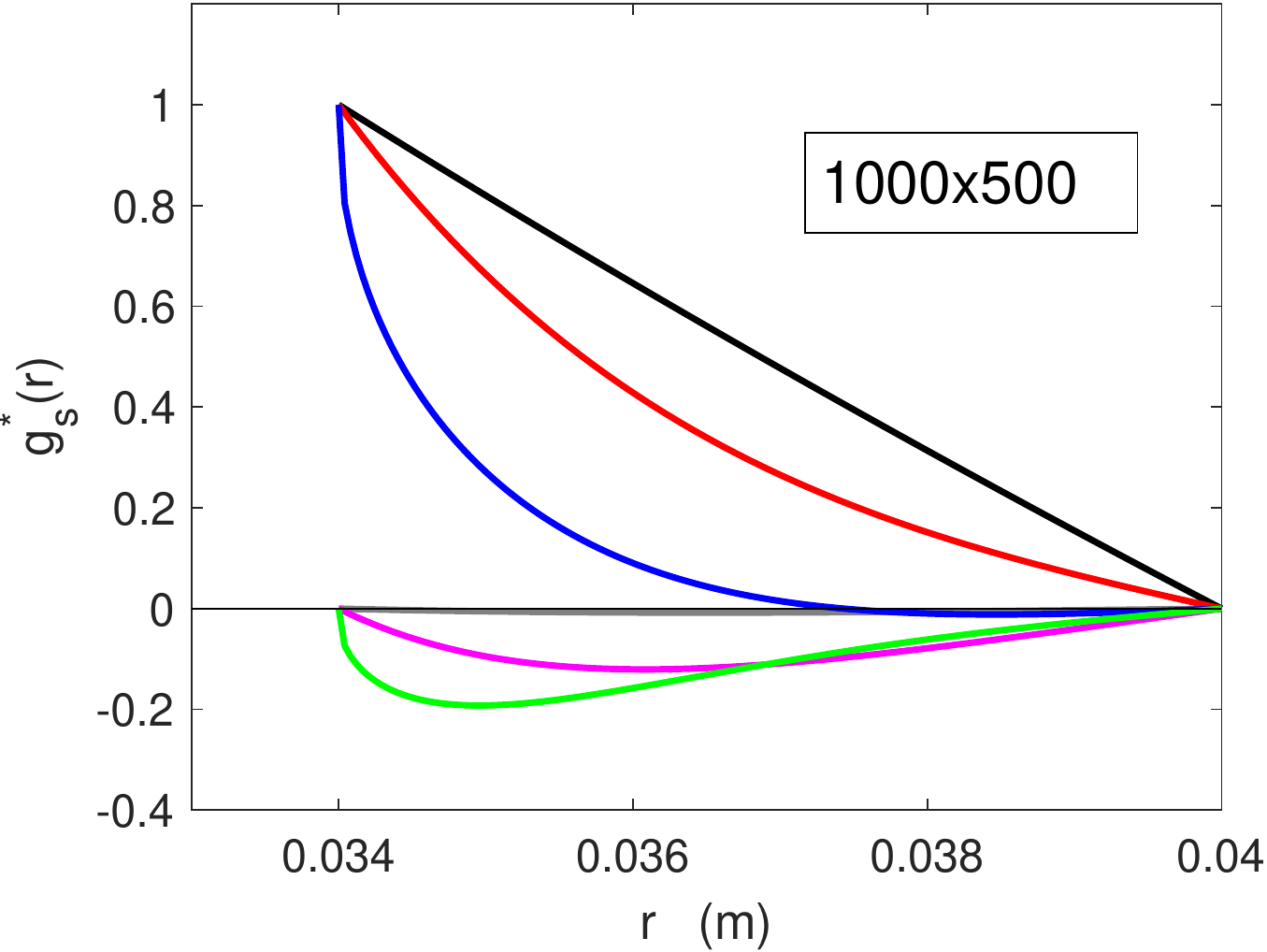}
  \end{minipage}
  \begin{minipage}{0.45\textwidth}
  \centering
  \includegraphics[width=\linewidth]{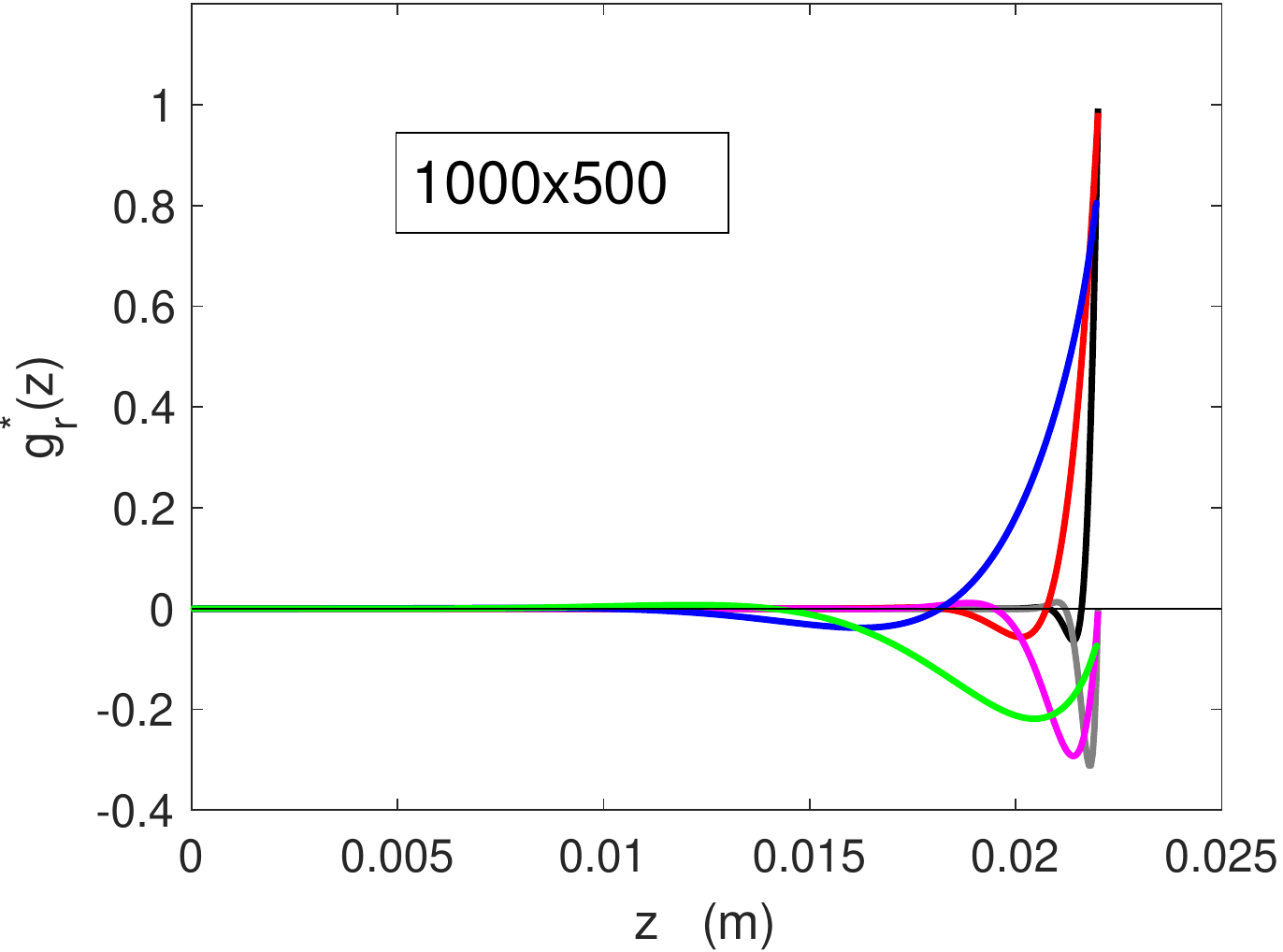}
  \end{minipage} 
  
  \begin{minipage}{0.45\textwidth}
  \centering
  \includegraphics[width=\linewidth]{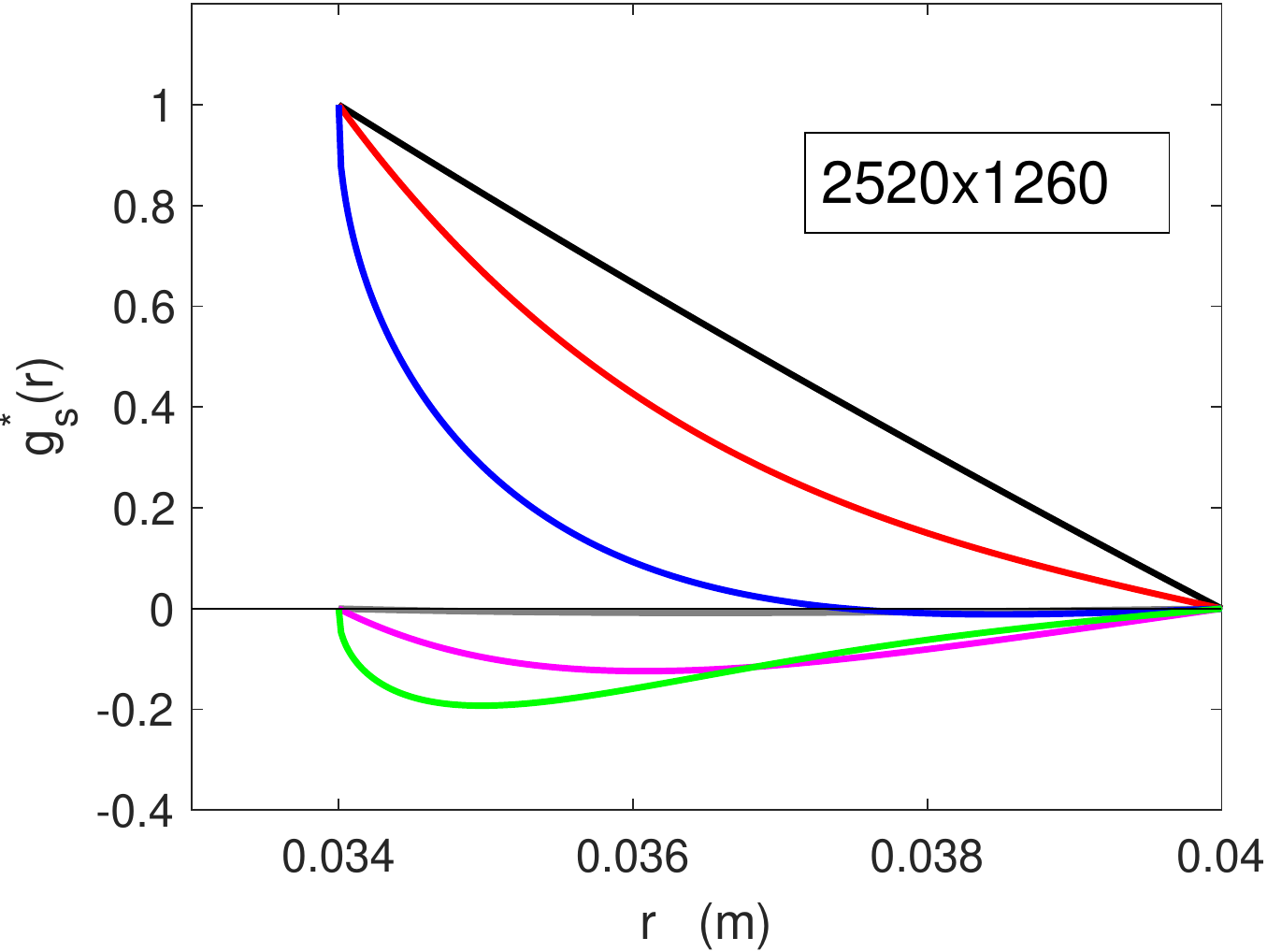}
  \end{minipage}
  \begin{minipage}{0.45\textwidth}
  \centering
  \includegraphics[width=\linewidth]{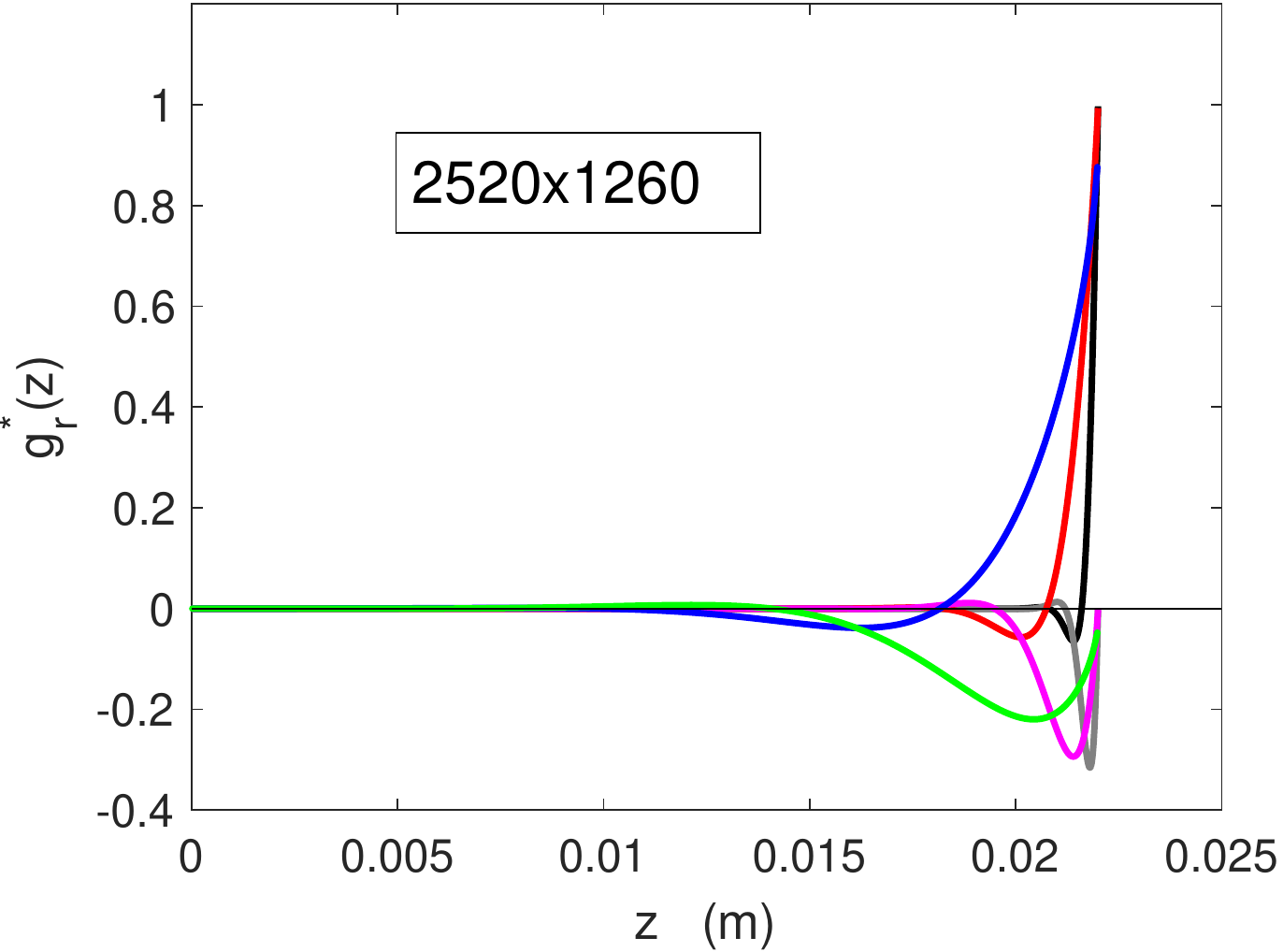}
  \end{minipage} 
\caption{Real and imaginary parts of the azimuthal velocity profiles at the interface (left) and the bicone rim vertical (right) obtained for different frequency and $Bo^*$ values. Top row: $200\times100$ mesh. Second row: $440\times220$ mesh. Third row: $1000\times500$ mesh. Bottom row: $2520\times1260$. Case A: High frequency with a viscoelastic interface (black and gray lines for the real and imaginary parts, respectively). Case B: Medium frequency with purely viscous interface (red and magenta lines for the real and imaginary parts, respectively). Case C: Low frequency and clean air-water interface (blue and green lines for the real and imaginary parts, respectively).}
\label{fig:g(rz)NxM}	
\end{figure}

A more illustrative way to look at these results is through the azimuthal velocity profile at the interface, $g^*_s(r) = g^*(R_b\le r\le R_c, z=h)$, and at the vertical line starting at the bicone rim, $g^*_r(z) = g^*(R_b, 0 \le z \le h)$. 
In Fig. \ref{fig:g(rz)NxM} we show graphs of $g^*_s(r)$ and $g^*_r(z)$, represented at the left and right columns, respectively, obtained with different mesh sizes, namely, $200\times100$ (top row), $440\times220$ (second row), $1000\times500$ (third row) and $2520\times1260$ (bottom row). In each graph three different dynamical situations are considered. We name case A (black and gray lines for the real and imaginary parts, respectively) the case of a viscoelastic interface ($\eta_s^* = (1-i)\times 10^{-3} \;\text{N}\cdot \text{s/m}$) at high frequency ($f = 5$ Hz). Case B (red and magenta lines for the real and imaginary parts, respectively) pertains to a purely viscous interface ($\eta_s^* = 10^{-5} \;\text{N}\cdot \text{s/m}$) at an intermediate frequency ($f = 0.5$ Hz). Finally, case C (blue and green lines for the real and imaginary parts, respectively) refers to a clean air-water interface at low frequency ($f = 0.05$ Hz).

Several aspects deserve particular attention here. For instance, at high values of the Boussinesq number, even at moderately high frequency, $\Re[g^*_s(r)]$ is linear and $\Im[g^*_s(r)]$ is negligible. Conversely, at low values of the Boussinesq number, $\Re[g^*_s(r)]$ decays close to the bicone rim, while $\Im[g^*_s(r)]$ gets values comparable to those of $\Re[g^*_s(r)]$. On the other hand, $\Re[g^*_r(z)]$ and $\Im[g^*_r(r)]$ in all cases take negligible values in all the $z$ range but very close to the bicone rim, where strongly nonlinear velocity gradients appear at low interfacial viscosities, even at low frequencies (see Fig. \ref{fig:g(rz)NxM}). Consequently, the mesh spacing has to be considered with care: these nonlinear velocity gradients are better represented the thinner the mesh. 

However, the criteria for mesh selection must include not only precision but also computational time. In Table \ref{table:Times} we indicate the computational time, in seconds, employed to solve the matrix problem $solve\_NS\_bicono.m$ call, which takes most of the calculation time in a regular desktop computer with a 4 nuclei Pentium Intel Core i5-4460 processor and $16$ Gb of RAM memory.

\begin{table}[H]
\centering
\begin{tabular}{ |c|c|c|c| }
\hline
& & & \\
Mesh & Case A & Case B & Case C \\
& & & \\
\hline
& & & \\ 
$200\times100$ & 0.12 & 0.12 & 0.12 \\  
& & & \\ 
\hline
& & & \\
$440\times220$ & 0.73 & 0.72 & 0.77 \\ 
& & & \\
\hline
& & & \\ 
$1000\times500$ & 5.31 & 5.28 & 5.44 \\ 
& & & \\
\hline
& & & \\ 
$2520\times1260$ & 56.51 & 57.63 & 55.44 \\ 
& & & \\
\hline
\end{tabular}
\caption{Time cost (in seconds) of solving the matrix problem for several mesh sizes in a desktop PC with a Pentium Intel Core i5-4460 processor and $16$ Gb of RAM memory.}
\label{table:Times}
\end{table}

\begin{figure}[H]
\centering
\includegraphics[width=.7\linewidth]{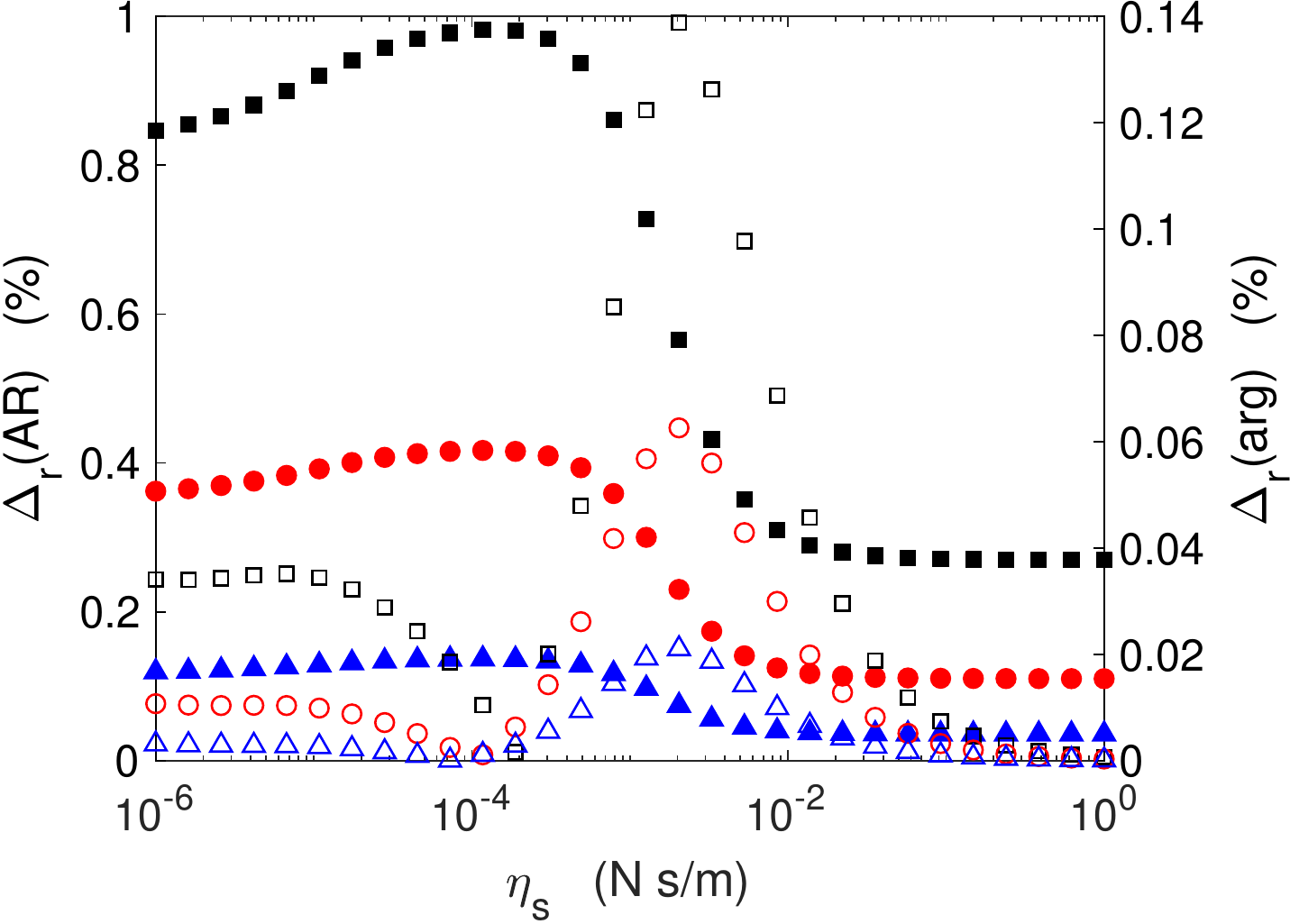}
\caption{Relative differences in modulus, $\Delta_r(AR)$ (solid symbols), and phase, $\Delta_r(arg)$ (open symbols), between the solutions obtained with different mesh sizes taking as a reference the $2520\times1260$ mesh solution. Black symbols: $200\times100$ mesh. Red symbols: $440\times220$ mesh. Blue symbols: $1000\times500$}
\label{fig:ARreldiff}
\end{figure}

To illustrate how the mesh size affects the complex amplitude ratio we have computed the relative differences in modulus and phase between the solutions obtained with the three mesh sizes of Fig. \ref{fig:g(rz)NxM} taking as a reference the solution for the $2520\times1260$ mesh size. More specifically, we have computed

\begin{align}
\Delta_r(AR) = \frac{\big|AR^*_{N\times M}\big| - \big|AR^*_{2520\times1260}\big|}{\big|AR^*_{2520\times1260}\big|},
\end{align}

\noindent and

\begin{align}
\Delta_r(\arg) = \Bigg| \frac{\arg (AR^*_{N\times M}) - \arg (AR^*_{2520\times1260})}{\arg (AR^*_{2520\times1260})} \Bigg|,
\end{align}

\noindent for purely viscous interfaces with $\eta_s$ in the range $10^{-6} \le \eta_s \le 1$ N$\cdot$s/m, and at the same frequency f = 0.5 Hz. The results are shown in Fig. \ref{fig:ARreldiff}. For the case of the $1000\times500$ mesh (blue symbols), the relative difference with the finest mesh is always below $0.2 \%$ in the modulus (solid symbols) and below $0.03 \%$ in the phase (open symbols). The $1000\times500$ mesh represents, therefore, a good compromise between resolution and computational cost and, consequently, we will use this mesh throughout the rest of this report.

\subsection{Consistency}

A direct test to check the consistency of the program results can be set up through the following two step strategy: i) Using the subroutine that solves the Navier-Stokes equation, synthesize the flow field for an interface with prescribed viscoelastic properties and obtain the corresponding modulus and argument of the complex torque/angle amplitude ratio, $AR^*$, and ii) use the obtained values for the modulus and argument of $AR^*$ as input data for the program and check whether or not the initially prescribed interfacial viscoelasticity is recovered. 

\begin{figure}[H]
  \begin{minipage}{0.5\textwidth}
  \centering
  \includegraphics[width=\linewidth]{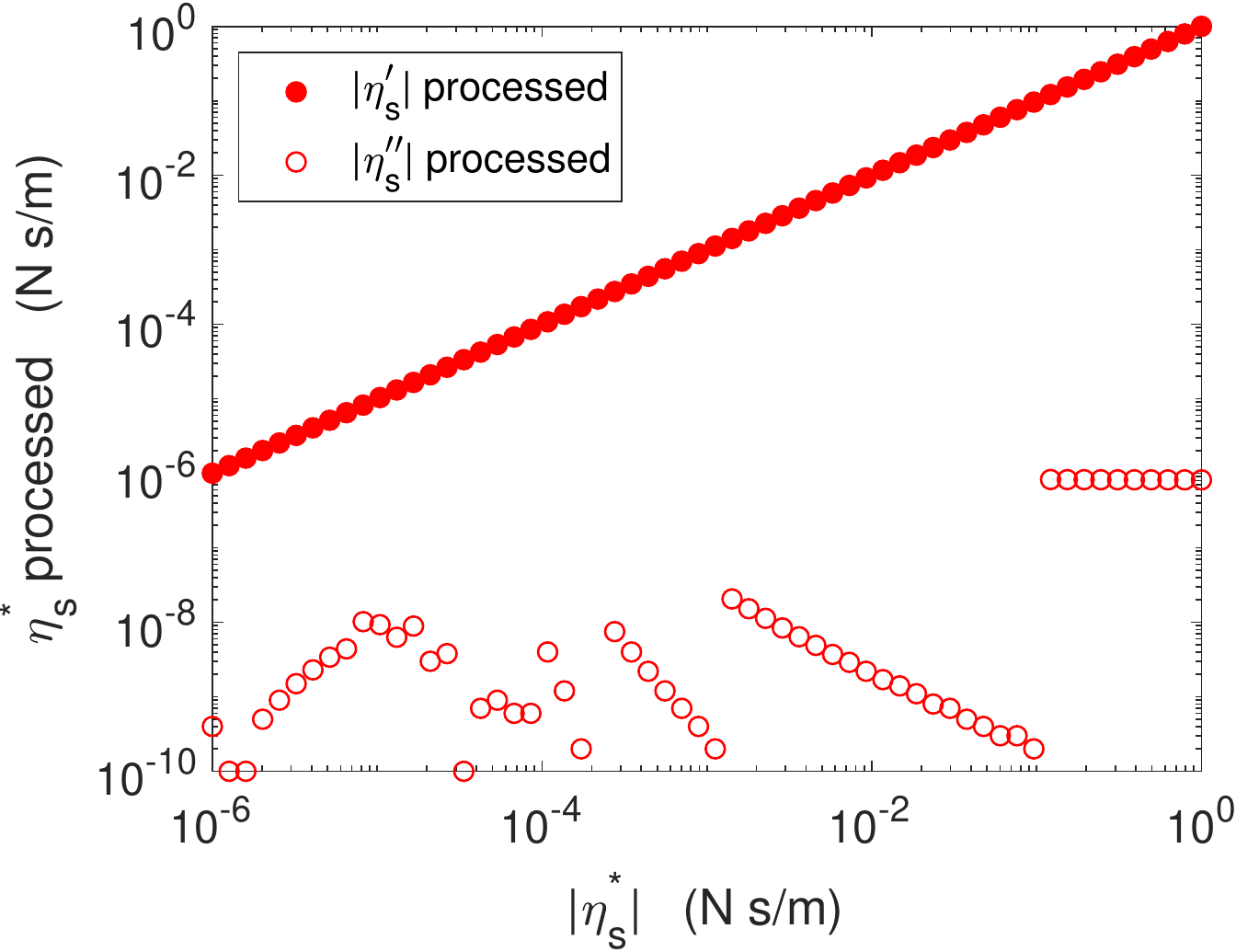}
  \end{minipage}
  \begin{minipage}{0.5\textwidth}
  \centering
  \includegraphics[width=\linewidth]{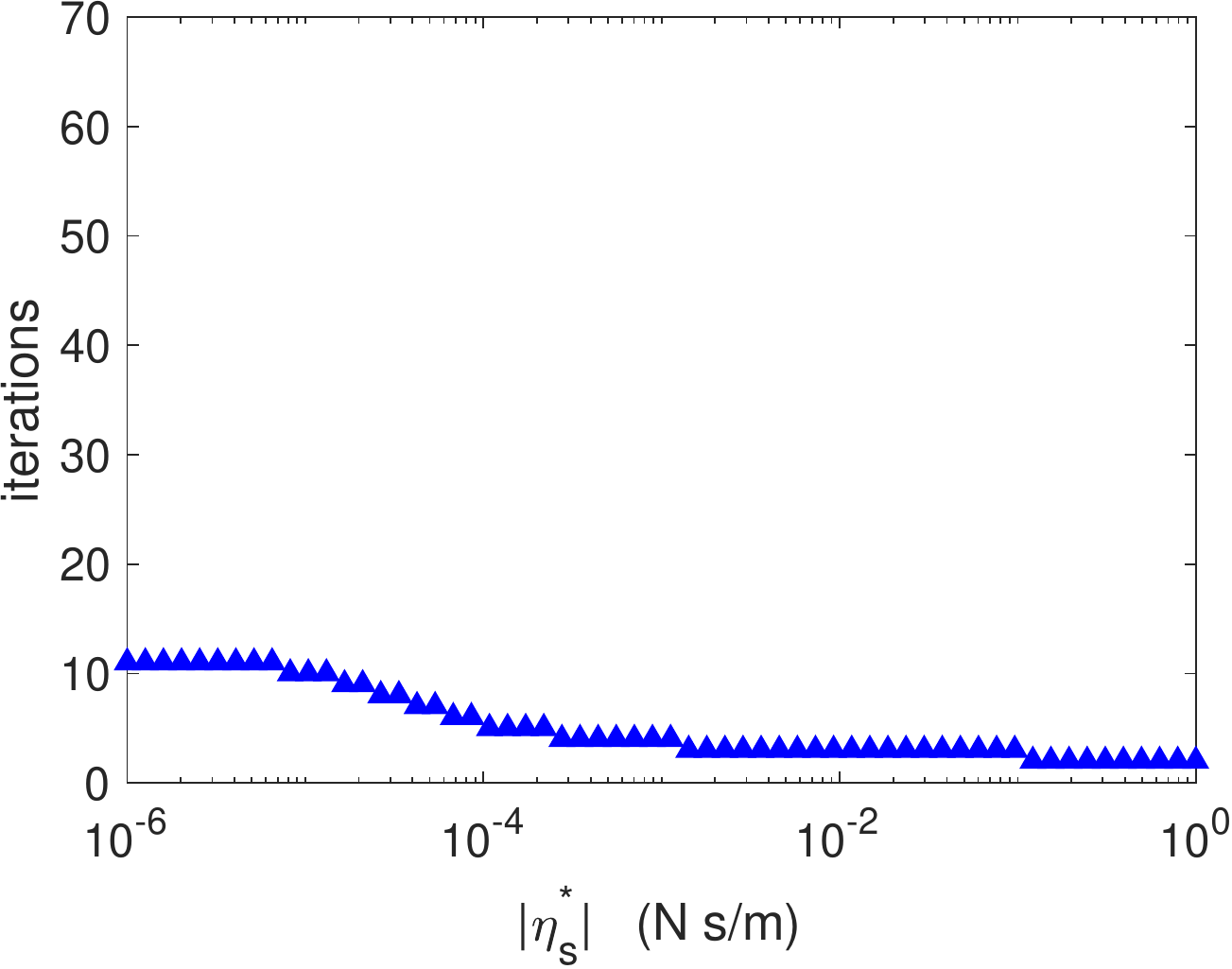}
  \end{minipage}
  
  \begin{minipage}{0.5\textwidth}
  \centering
  \includegraphics[width=\linewidth]{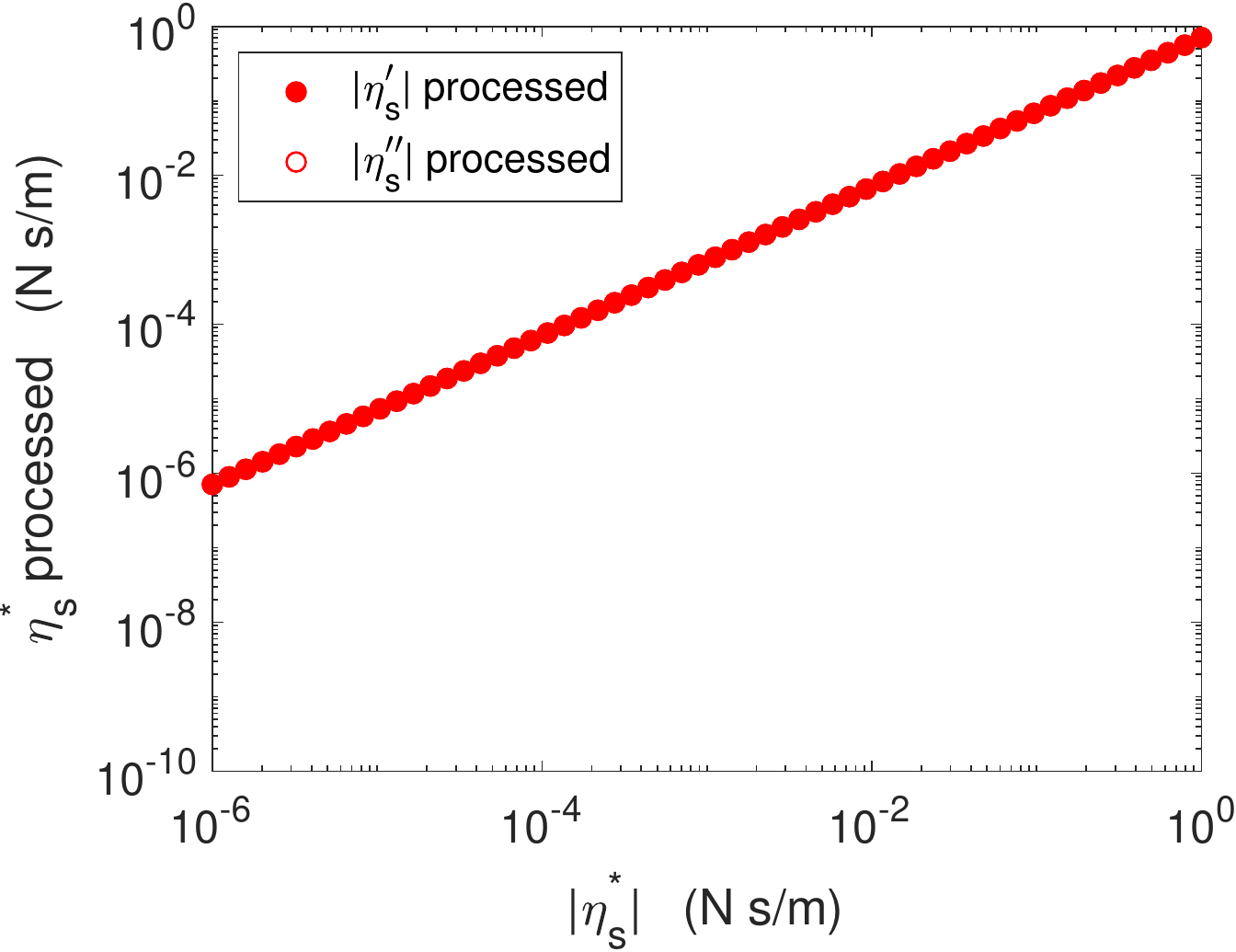}
  \end{minipage}
  \begin{minipage}{0.5\textwidth}
  \centering
  \includegraphics[width=\linewidth]{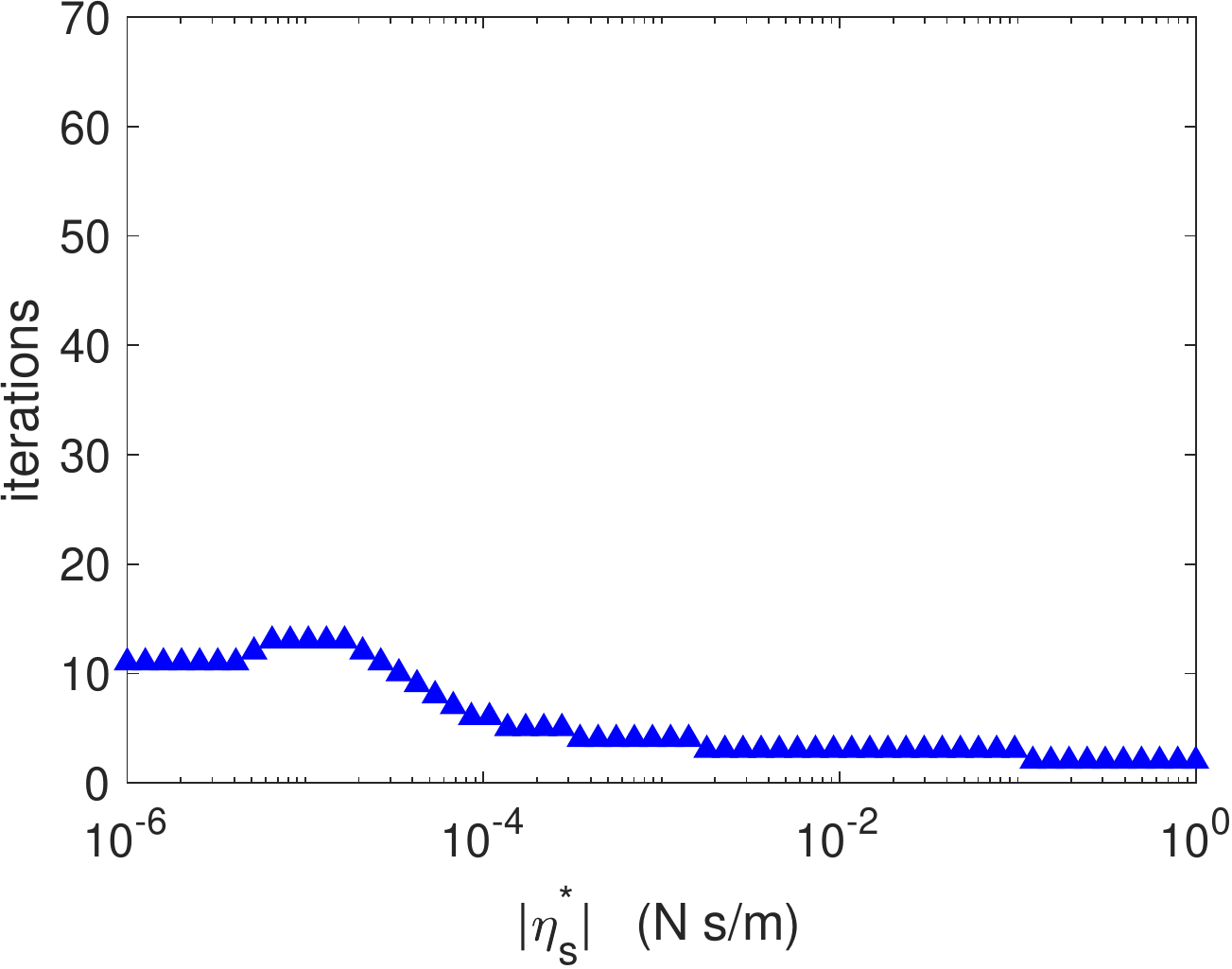}
  \end{minipage}

  \begin{minipage}{0.5\textwidth}
  \centering
  \includegraphics[width=\linewidth]{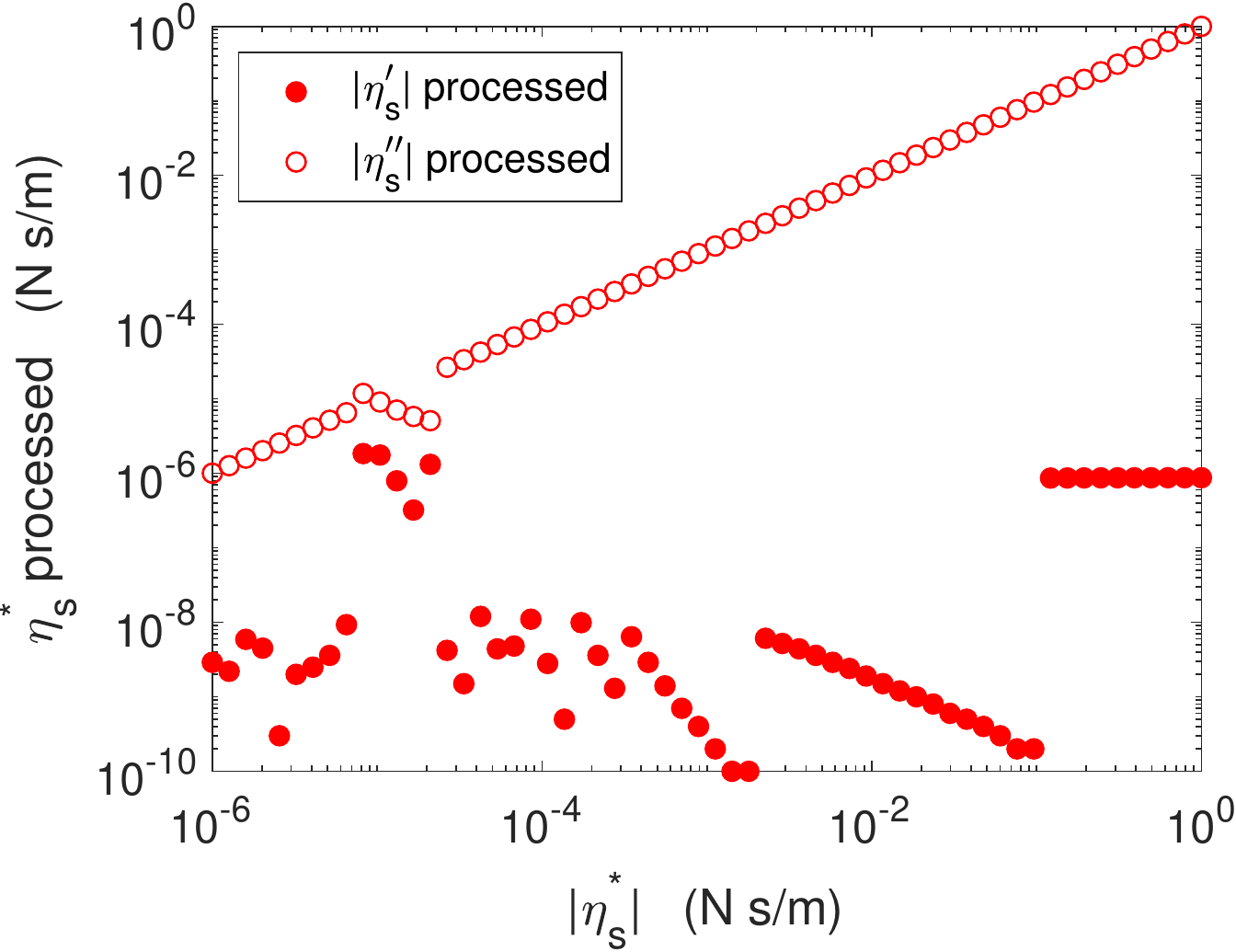}
  \end{minipage}
  \begin{minipage}{0.5\textwidth}
  \centering
  \includegraphics[width=\linewidth]{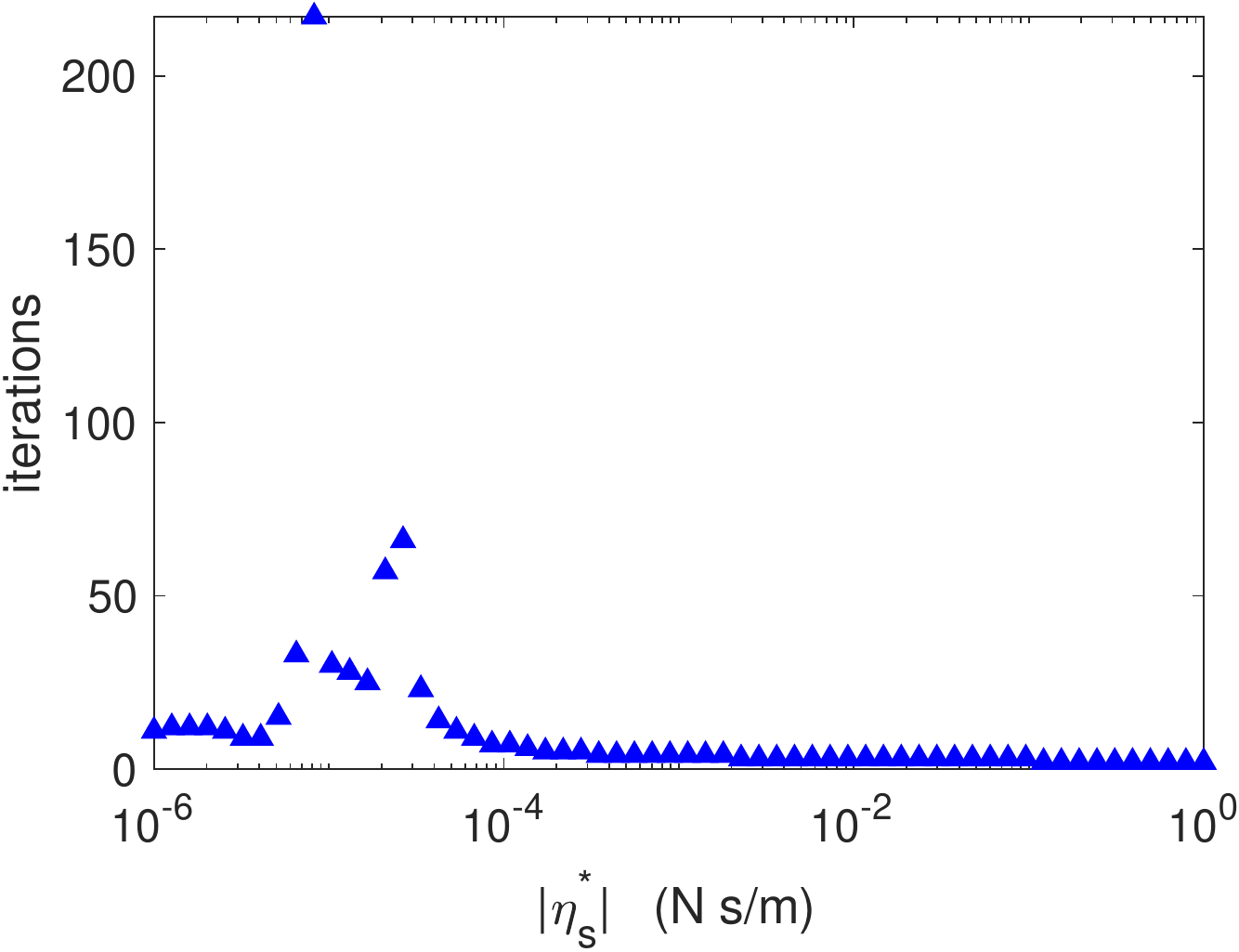}
  \end{minipage}

\caption{Real (triangles) and imaginary (circles) parts of the interfacial shear viscosity (left column) and number of iterations until convergence (right column) as a function of the interfacial shear viscosity modulus, $\vert \eta^*_s \vert$ at a frequency $f = 0.5$ Hz. Upper row: Purely viscous interface ($\eta_s''=0$). Middle row: Viscoelastic interface ($\eta'_s = \eta''_s$). Bottom row: Purely elastic interface ($\eta_s'=0$). }
\label{fig:Eta_sSweep}	
\end{figure}

In Fig. \ref{fig:Eta_sSweep} we show the results of such a procedure in the cases of a) a purely viscous interface, $\eta_s^* = \eta_s'$, b) a viscoelastic interface whose complex viscosity has equal real and imaginary parts, i.e., $\eta^*_s = \eta_s' - i \eta_s''$, and c) a purely elastic interface, $\eta_s^* = - i\eta_s''$. In all three cases, the modulus of the complex interfacial viscosity $|\eta_s^*|$ has the same value. The results shown here correspond to calculations performed with $N = 1000$, $M = 500$ and $tolMin = 10^{-5}$. In all of the dynamical calculations we have used the value of the rheometer frictional torque parameter, $b =3.2\times 10^{-8}\mbox{\,kg$\cdot$m$^{2}\cdot$rad/s}$, which corresponds to our Bohlin-CVOR rheometer \cite{Tajuelo2018}.

The graphs at the left column of Fig. \ref{fig:Eta_sSweep} represent the values obtained for the real and imaginary parts of $\eta_s^*$ as a function of the programmed $\eta_s^*$ value. The graphs at the right column represent the number of iterations needed for convergence in each case. 

Nice agreement between the programmed and obtained values for $\eta_s^*$ are found. In the cases of the purely viscous ($\eta_s''=0$) and the purely elastic ($\eta_s'=0$) interfaces the values obtained for $|\eta_s'|$ and $|\eta_s''|$, respectively, fairly coincide with the programmed values with the exception of five data points that deviate in the case of a purely elastic interface (bottom left graph in Fig. \ref{fig:Eta_sSweep}). In the case of a viscoelastic interface (middle left graph in Fig. \ref{fig:Eta_sSweep}) the values of $\eta_s'$ and $\eta_s''$ perfectly overlap in the log-log plot. 

Small positive or negative non null values of $\eta'_s$ and $\eta''_s$ appear for the purely viscous and the purely elastic interfaces that are caused by numerical errors due to the finite tolerance. In the case of the purely viscous interface these non null erroneous values are at least two orders of magnitude below those corresponding to the non null programmed values. Furthermore, this difference increases if lower values of $tolMin$ are used. In the case of the purely elastic interface, larger errors in the supposedly null value of $\eta'_s$ appear, particularly in the above mentioned points that deviate from the programmed $\eta''_s$ value.

As shown in the plots at the right column in Fig. \ref{fig:Eta_sSweep}, in all the cases here considered convergence occurred most often in about $10$ iterations. Remarkably, in the case of the purely elastic interface five values points need a larger number of iterations with one of them needing more than $200$ iterations for convergence to occur. Preliminary work suggests that this peak might be related to a resonance phenomenon probably similar to the well-known resonance problem appearing in rotational rheometry of weak gels \cite{Baravian2007}. Clarifying this point is, however, beyond the scope of this report.

\subsection{Performance tests on experimental data}

Finally, we check the performance of the program by processing real experiments on known viscosity newtonian interfacial films and fatty acids Langmuir monolayers. All the computations reported in this section have been made with $N = 1000$, $M = 500$, and $tolMin = 10^{-5}$.

Thin films of newtonian liquids non-miscible with the subphase can be used to construct newtonian interfaces with well known interfacial viscosity. Indeed, the effective interfacial viscosity of a film, with depth $d$, of a newtonian fluid, with bulk viscosity $\eta$, is given by the expression $\eta_s = d\cdot\eta$. This fact may be conveniently used to benchmark the performance of the program by constructing newtonian interfaces with tailored interfacial viscosity.

\begin{figure}[H]
\centering
\includegraphics[width=.7\linewidth]{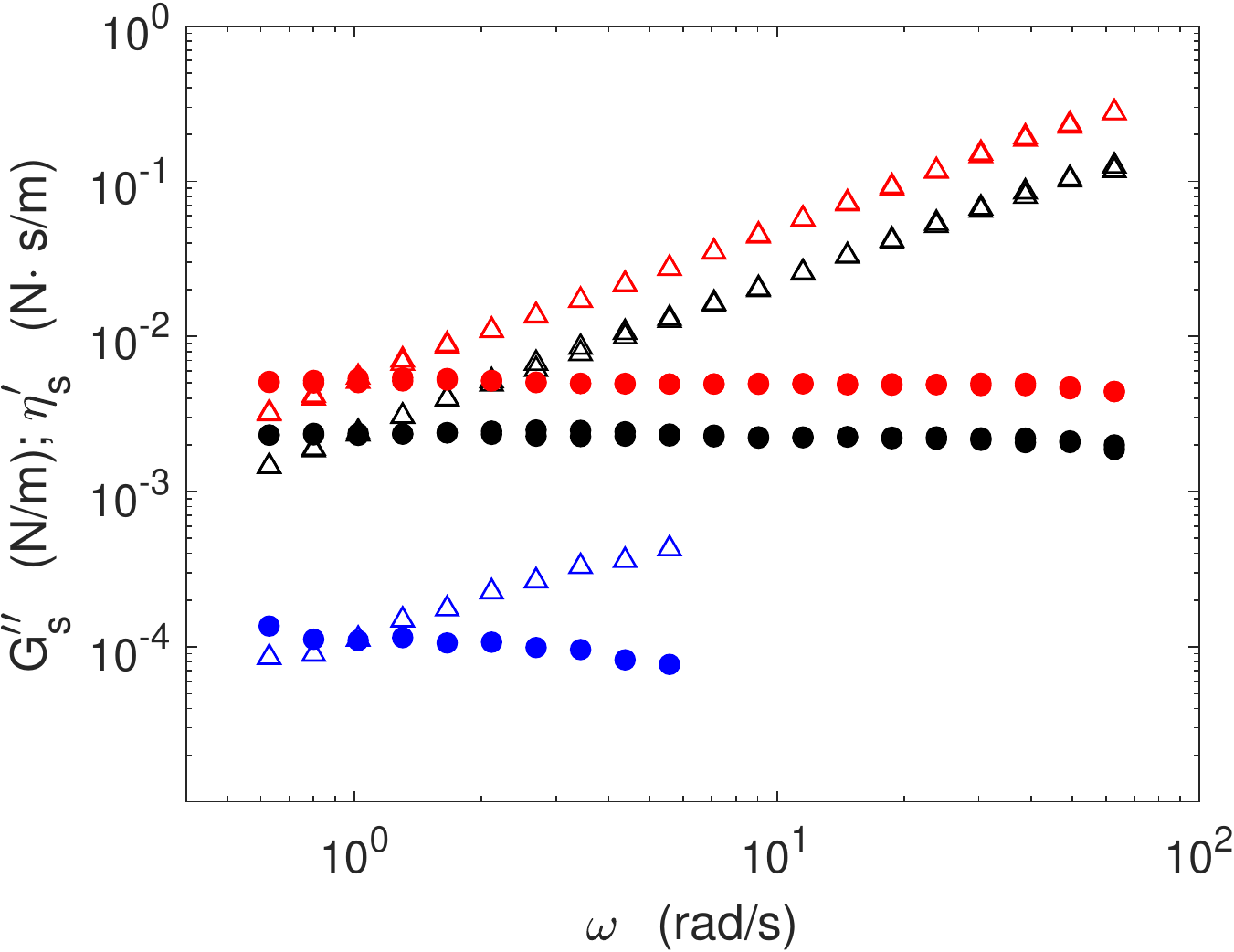}
\caption{Interfacial loss modulus, $G_s''$ (circles), and interfacial viscosity, $\eta_s'$ (triangles), of the thin films of silicone oil described in the text.}
\label{fig:SiOilfsweep}
\end{figure}

In Fig. \ref{fig:SiOilfsweep} we show the results of processing experimental data obtained at an air/water interface covered with three different silicone oil films. Red symbols: $\eta= 33$ Pa$\cdot$s, $d=150\; \mu$m; $\eta_s= 5\times 10^{-3}$ N$\cdot$s/m. Black symbols: $\eta= 33$ Pa$\cdot$s, $d=75\; \mu$m; $\eta_s= 2.5\times 10^{-3}$ N$\cdot$s/m. Blue symbols: $\eta= 1$ Pa$\cdot$s, $d=100\; \mu$m; $\eta_s= 10^{-4}$ N$\cdot$s/m. The values obtained for the interfacial viscosity agree very well with the expected value in the three cases. Furthermore, the loss modulus grows linearly with frequency with a slope equal to 1 in the doubly logarithmic plot. In all of the three cases convergence occurred in 5 iteration steps or less.

\begin{figure}[H]
\centering
\includegraphics[width=.7\linewidth]{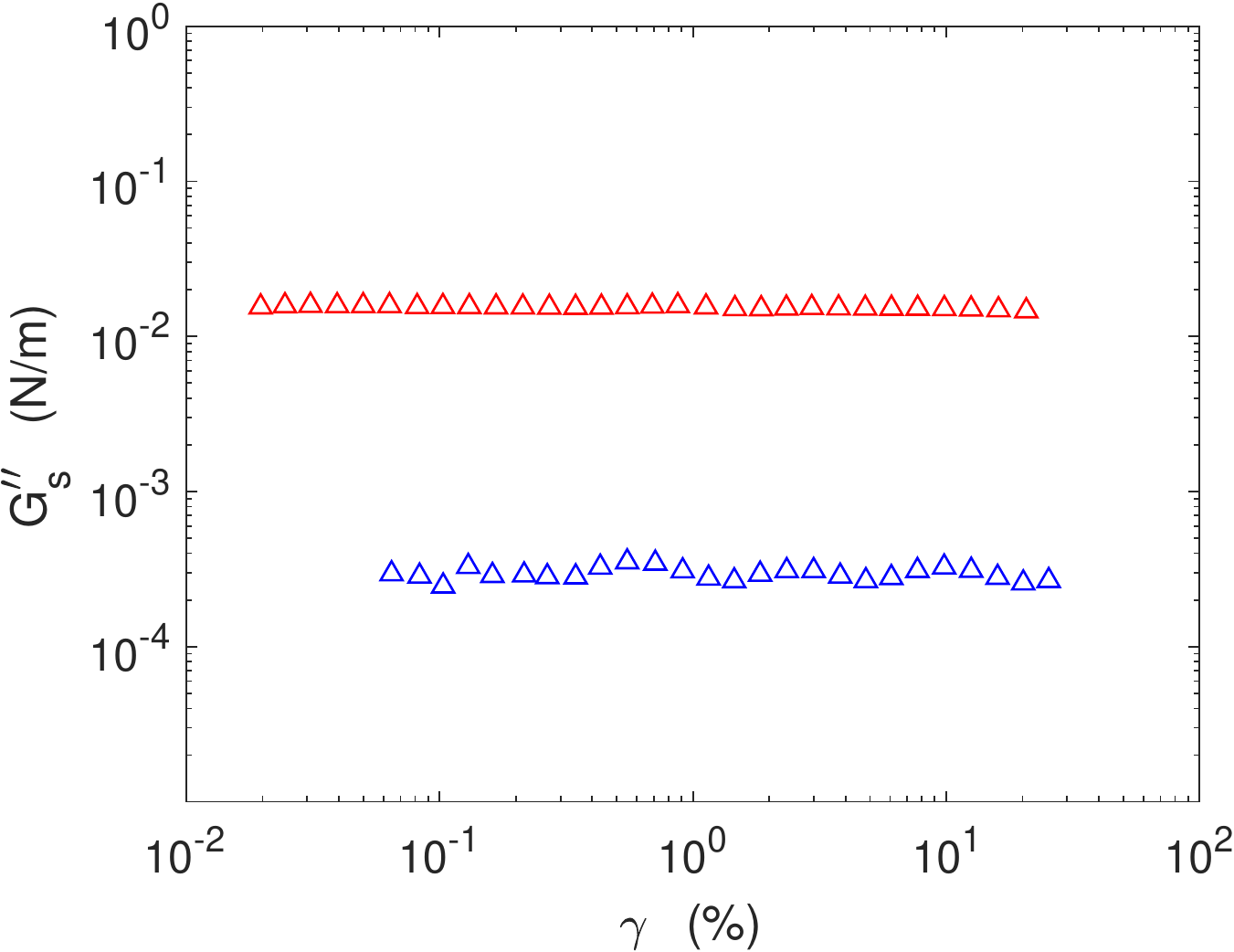}
\caption{Interfacial loss modulus, $G_s''$ as a function of strain, $\gamma$, for the silicone oil films with $\eta_s= 5\times 10^{-3}$ N$\cdot$s/m (red circles) and $\eta_s= 10^{-4}$ N$\cdot$s/m (blue circles) interfacial viscosity.}
\label{fig:SiOilasweep}
\end{figure}

We also show in Fig. \ref{fig:SiOilasweep} plots of the values obtained for the loss modulus $G''$ at a fixed frequency $f = 0.5$ Hz ($\omega = \pi$ rad/s) when varying the oscillation amplitude. They show that all of these measurements have been made within the linear regime.

\begin{figure}[H]
\centering
\includegraphics[width=.7\linewidth]{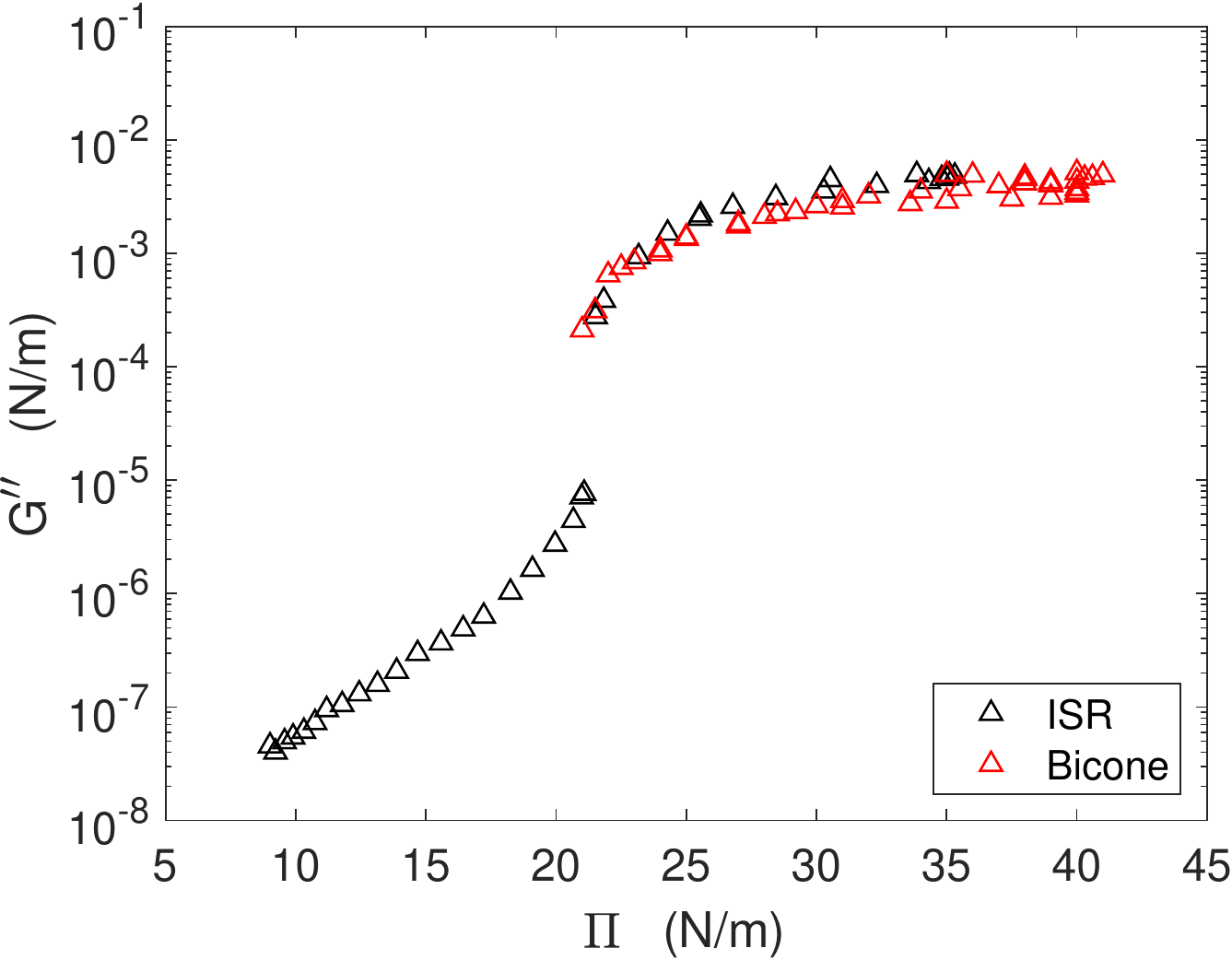}
\caption{Comparison of the measurements of the loss modulus of a Pentadecanoic acid Langmuir monolayer obtained with the bicone rheometer (red dots) and the magnetic tweezers ISR (black dots).}
\label{fig:PDA}
\end{figure}

In Fig. \ref{fig:PDA} we show the results obtained by processing the bicone rheometer data (red dots) obtained for a pentadecanoic acid Langmuir monolayer on an air/water interface. For comparison we also show the corresponding results (black dots) obtained by means of a magnetic tweezers interfacial shear rheometer \cite{Tajuelo2016,Tajuelo2017}. An isothermal compression was applied to the monolayer so that a transition from the L2 phase to the LS phase occurred. The storage modulus is not shown because at the L2 phase it is too small to be measured with the bicone rheometer and at the LS phase it is too small to be measured with any of the two rheometers \cite{Tajuelo2017}. The data obtained for the loss modulus with the two rheometers show an excellent agreement. Here again, convergence occurred in 6 iteration steps or less.

\section*{Acknowledgements}

P.S.P acknowledges D. Chaos for fruitful advice on Matlab programming. The authors acknowledge partial support from MINECO (Grants No. FIS2013-47350-C5-5-R and No. FIS2017-86007-C3-3-P), J.M.P. from MINECO (Grant No. MTM2015-63914-P). J.T. acknowledges a grant from the UNED's Researchers Formation Program, and P.S.P. acknowledges a contract funded by Comunidad de Madrid (ref. PEJ16/IND/AI-1253).

\hspace{2cm}
\appendix

\section{Example of starting script}
\label{section:exampleScript}

\begin{verbatim}
%This is an example of a script that calls 
% Postprocessing_Bicone_CPC.m
close all
clear,clc
% Geometry parameters
h=0.022;                      % distance between interface and
                               cup bottom [m]
R1=0.0340;                    % bicone radius [m]
R=0.04;                       % cup radius [m]
% Parameters of the rheometer dynamics
inertia=0.0000242019;         % system (rotor + bicone) inertia [Kg.m^2]
b=3.2e-8;                     % frictional torque contribution
                               [Kg.m^2.rad/s]
% Mesh parameters
N=200;                        % Subintervals in r direction
M=100;                         % Subintervals in z direction
% Subphase physical parameters
rho_bulk=1000;                % Subphase density [Kg/m^3]
eta_bulk=1e-3;                % Complex subphase viscosity [Pa.s]
% Iterative scheme parameters
iteMax=100;                   % maximum number of iterations
tolMin=0.00001;               % threshold tolerance
% Input/output data
colIndexAR=2;                 % ordinal number of the data of
                               the column that contains the
                               modulus of the amplitude ratio
colIndexDelta=3;              % ordinal number of the data of
                               the column that contains the
                               modulus of the amplitude ratio
colIndexFreq=1;               % ordinal number of the data of
                               the column that contains the
                               modulus of the amplitude ratio
inputFilepath=pwd;        % input filepath
outputFilepath=pwd;       % output filepath

% Execute postprocessingBiconeCPC.m with the specified input data
[GData,etasData,bouData,ARcalcData,deltaARcalcData,iterationsTimesData,
iterationsData,timeElapsedTotal]=postprocessingBiconeCPC(h,R1,R,inertia,
b,N,M,rho_bulk,eta_bulk,iteMax,tolMin,colIndexAR,colIndexDelta,colIndexFreq,
inputFilepath,outputFilepath);
\end{verbatim}

\section{Details of the numerical scheme}
\label{Apendix_Script}

\subsection{Filling the coefficients matrix and the independent terms vector}
\label{FD-details}
The mesh points in the $(\bar{r},\bar{z})$ coordinates are indexed as $(j,k)$. However, the origin in the $(\bar{r},\bar{z})$ representation is located at the lower left corner while the origin in the $(j,k)$ representation is placed at the upper left corner. Hence, the $k$ index and the coordinate $\bar{z}$ have opposite senses.
We can express $g^*(\bar{r},\bar{z})$ values on nodes as follows
\begin{align}
g_{j,k}^*=g^*\left( (j-1)\frac{1}{N},(k-1)\frac{\bar{h}}{M} \right),\hspace{0.5cm}
\forall j,k\in\mathbb{Z}\, / \, 1\leq j\leq N+1\, , \, 1\leq k\leq M+1.
\end{align}

The expressions of the discretized second order partial derivatives on the mesh nodes are:
\begin{align}
&\left( \frac{\partial^2 g_{j,k}^*}{\partial r^2} \right)_{j=j',k=k'}=N^2\left(g_{j'+1,k'}^*-2g_{j',k'}^*+g_{j'-1,k'}^*\right),\nonumber\\
&\left( \frac{\partial^2 g_{j,k}^*}{\partial z^2} \right)_{j=j',k=k'}=\left(\frac{M}{\bar{h}}\right)^2\left(g_{j',k'-1}^*-2g_{j',k'}^*+g_{j',k'+1}^*\right).
\end{align}

In the following we describe how the $\textbf{A}$ matrix is filled. More precisely, we give the expressions for the values of the matrix elements describing the internal nodes, the Boussinesq-Scriven boundary condition at the interface, the no-slip boundary conditions and symmetry condition. 

\begin{figure}[H]
\centering
\includegraphics[scale=1]{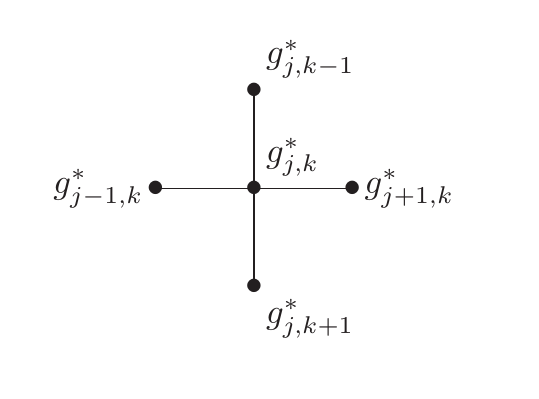}
\caption{Five nodes scheme}
\label{fig:5nodes}
\end{figure}

For the internal nodes (nodes not at a boundary) a five nodes formula, based on the scheme illustrated in Fig. \ref{fig:5nodes}, can be obtained for the Navier-Stokes equation having the following form:
\begin{align}
i\,Re\,g_{j,k}^*=&N^2\left( g_{j+1,k}^*+g_{j-1,k}^*-2g_{j,k}^*+\frac{g_{j+1,k}^*-g_{j-1,k}^*}{2(j-1)}-\frac{g_{j,k}^*}{(j-1)^2} \right)\nonumber\\
&+\left( \frac{M}{\bar{h}} \right)^2\left( g_{j,k-1}^*+g_{j,k+1}^*-2g_{j,k}^* \right),\nonumber\\
&\forall j,k\in\mathbb{Z}\, / \, 2\leq j\leq N,\,2\leq k\leq M.\label{ap2-navier-nodos}
\end{align}

Factorizing and rearranging the terms in $g_{j,k}^*$ we obtain the following expression:
\begin{align}
g_{j,k-1}^*\left[\left(\frac{M}{\bar{h}}\right)^2\right]&+g_{j-1,k}^*\left[N^2\left(1-\frac{1}{2(j-1}\right)\right] \nonumber  \\ 
&+g_{j,k}^*\left[-iRe-N^2\left(2+\frac{1}{(j-1)^2}\right)-2\left(\frac{M}{\bar{h}}\right)^2\right] \nonumber \\
&+g_{j+1,k}^*\left[N^2\left(1+\frac{1}{2(j-1)}\right)\right]+g_{j,k+1}^*\left[\left(\frac{M}{\bar{h}}\right)^2\right]=0, \nonumber \\
&\forall j,k\in\mathbb{Z}\, /\hspace{0.5cm} \, 1\leq j\leq N-1,\hspace{0.5cm}1\leq k\leq M-1
\end{align}

Then the elements of the coefficient matrix for the internal points are:
\begin{align}
&A((k-1)*(N+1)+j,\hspace{0.5cm}((k-2)*(N+1)+j)=\left(\frac{M}{\bar{h}}\right)^2  \nonumber \\
&A((k-1)*(N+1)+j,\hspace{0.5cm}((k-1)*(N+1)+j-1)=N^2\left(1-\frac{1}{2(j-1)}\right)\nonumber \\
&A((k-1)*(N+1)+j,\hspace{0.5cm}((k-1)*(N+1)+j)=\nonumber \\
& \qquad \qquad \qquad \qquad \qquad \qquad -iRe-N^2\left(2+\frac{1}{(j-1)^2}\right)-2\left(\frac{M}{\bar{h}}\right)^2\nonumber \\
&A((k-1)*(N+1)+j,\hspace{0.5cm}((k-1)*(N+1)+j+1)=N^2\left(1+\frac{1}{2(j-1)}\right)\nonumber \\
&A((k-1)*(N+1)+j,\hspace{0.5cm}(k*(N+1)+j)=\left(\frac{M}{\bar{h}}\right)^2\nonumber \\
&\forall j,k\in\mathbb{Z}\, /\hspace{0.5cm} \, 2\leq j\leq N,\hspace{0.5cm}2\leq k\leq M
\end{align}

The Boussinesq-Scriven boundary condition (Eq. \ref{boundary-bouss}) is written as
\begin{align}
\frac{M}{\bar{h}}\left( g_{j,1}^*-g_{j,2}^* \right)=Bo^*N^2&\left( \frac{g_{j+1,1}^*-g_{j-1,1}^*}{2(j-1)}-\frac{g_{j,1}^*}{(j-1)^2}+g_{j-1,1}^*+g_{j+1,1}^*-2g_{j,1}^* \right),\nonumber\\
\forall j&\in\mathbb{Z}\, / \, \lfloor N\bar{R_b}\rfloor+1<j<N+1.
\label{Apendix-Bouss}
\end{align}

\begin{figure}[H]
\centering
\includegraphics[scale=1]{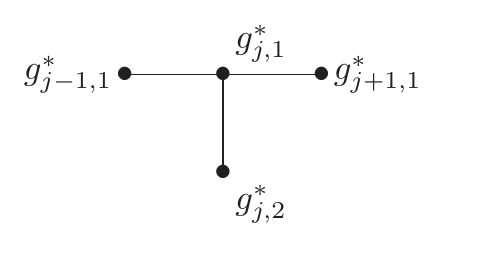}
\caption{Four nodes scheme}
\label{fig:4nodes}
\end{figure}

Implementing the corresponding four node formula (see Fig. \ref{fig:4nodes}), factorizing and rearranging the terms in $g_{j,k}^*$ in the Boussinesq-Scriven condition, Eq. \ref{Apendix-Bouss} we arrive at

\begin{align}
& g_{j-1,1}^*\left[Bo^*N^2\left(1-\frac{1}{2(j-1)}\right)\right]& \nonumber \\
&+g_{j,1}^*\left[-Bo^*N^2\left(2+\frac{1}{(j-1)^2}\right)-\frac{M}{\bar{h}}\right] \nonumber \\
&+g_{j+1,1}^*\left[Bo^*N^2\left(1+\frac{1}{2(j-1)}\right)\right]+g_{j,2}^*\left[\left(\frac{M}{\bar{h}}\right)\right] = 0 \nonumber \\
&\forall j\in\mathbb{Z}\, / \, \lfloor N\bar{R_b}\rfloor+1<j<N+1.
\end{align}

Hence, the corresponding elements in the matrix are:
\begin{align}
&A(j,\hspace{0.5cm}j-1)=Bo^*N^2\left(1-\frac{1}{2(j-1)}\right)\nonumber \\
&A(j,\hspace{0.5cm}j)=-Bo^*N^2\left(2+\frac{1}{(j-1)^2}\right)-\frac{M}{\bar{h}}\nonumber \\
&A(j,\hspace{0.5cm}j+1)=Bo^*N^2\left(1+\frac{1}{2(j-1)}\right)\nonumber \\
&A(j,\hspace{0.5cm}j+(N+1))=\left(\frac{M}{\bar{h}}\right)\nonumber \\
&\forall j\in\mathbb{Z}\, / \, \lfloor N\bar{R_b}\rfloor+2\leq j\leq N.
\end{align}

The no-slip and symmetry boundary conditions in Eq. \ref{boundary1} are then written as \footnote{$\lfloor x \rfloor$ indicates the highest integer smaller than or equal to $x$.}
\begin{align}
&g_{j,M+1}^*=0,&&\forall j\in\mathbb{Z}\, / \, 1\leq j\leq N+1,\\
&g_{N+1,k}^*=0,&&\forall k\in\mathbb{Z}\, / \, 1\leq k\leq M,\\
&g_{1,k}^*=0,&&\forall k\in\mathbb{Z}\, / \, 2\leq k\leq M,\\
&g_{j,1}^*=\frac{(j-1)}{N\bar{R_b}},&&\forall j\in\mathbb{Z}\, / \, 1\leq j\leq\lfloor N\bar{R_b}\rfloor+1.
\label{Eq:bvels}
\end{align}

These four conditions are implemented by means of their respective 1 values at the diagonal of $\textbf{A}$ as follows:
\begin{align}
&A(j+M*(N+1),\hspace{0.5cm}j+M*(N+1))=1,&&\forall j\in\mathbb{Z}\, / \, 1\leq j\leq N+1,\nonumber \\
&A(j*(N+1),\hspace{0.5cm}j*(N+1))=1,&&\forall k\in\mathbb{Z}\, / \, 1\leq k\leq M,\nonumber \\
&A(j*(N+1),\hspace{0.5cm}j*(N+1))=1,&&\forall k\in\mathbb{Z}\, / \, 2\leq k\leq M,\nonumber \\
&A(j,\hspace{0.5cm}j)=1,&&\forall j\in\mathbb{Z}\, / \, 1\leq j\leq\lfloor N\bar{R_b}\rfloor+1.
\end{align}

In order to fully exploit the MATLAB sparse matrix managing routines, three separate arrays containing the non-null values of the $\textbf{A}$ matrix elements and their corresponding row and column indexes are created. Subsequently, these arrays are used to construct an sparse matrix of size $(N+1)(M+1)\times(N+1)(M+1)$. We define vector $\textbf{b}$ also as a sparse column array with size $1\times (N+1)(M+1)$ whose non-null values are assigned according to Eq. \ref{Eq:bvels}. 

\subsection{Computing the subphase drag integral}
Finally, the subphase drag is computed using the compound trapezium rule. More explicitly, to obtain the complex amplitude ratio $AR^*$ it is mandatory to calculate the following integral:
\begin{align}
I=\int_0^{R_b}r^2\left(\frac{\partial g^*}{\partial z}\right)\Big|_{z=h}dr,
\end{align}
\noindent which, due to the spatial discretization over the bicone surface, is calculated actually as:
\begin{align}
I\simeq\int_0^{R_c\frac{N_b}{N}}r^2\left(\frac{\partial g^*}{\partial z}\right)\Big|_{z=h}dr,
\end{align}
\noindent that, using the compound trapezium rule, can be expressed as:

\begin{align}
I & \sim \frac{(\Delta r)^3}{2\Delta z}\left[ \sum_{i=2}^{N_b}(i-1)^{2}2(g^*(i)-g^*(i+(N+1)))\right. \nonumber \\
& \qquad + \left. N_b^2(g^*(N_b+1)-g^*((N_b+1)+(N+1)))\right]
\label{eq:numer_integr}
\end{align}



\bibliographystyle{elsarticle-num}



\end{document}